\documentclass[longauth]{aa}  

\usepackage{graphicx}
\usepackage{txfonts}
\usepackage{amssymb}
\usepackage{amsmath}
\usepackage{graphicx}
\usepackage[colorlinks=true, allcolors=blue]{hyperref}
\usepackage{booktabs}
\usepackage{amsfonts}     
\usepackage[T1]{fontenc} 
\usepackage[utf8]{inputenc} 
\usepackage{multicol}
\usepackage{xcolor}
\usepackage[switch]{lineno}
\usepackage{gensymb}
\usepackage{soul}
\usepackage{rotating}
\usepackage{comment}
\usepackage{multirow}
\usepackage{longtable}
\usepackage{lscape}
\usepackage{color}

\newcommand{\eqb}{\begin{eqnarray}}
\newcommand{\eqe}{\end{eqnarray}}

\begin{document}

\title{IXPE observation of PKS~2155$-$304 reveals the most highly polarized blazar}

\subtitle{}

\author{
Pouya M. Kouch \inst{\ref{UTU},\ref{FINCA},\ref{MRO}} \thanks{\href{mailto:pouya.kouch@utu.fi}{pouya.kouch@utu.fi}} \orcid{0000-0002-9328-2750}
Ioannis Liodakis \inst{\ref{NASA_Alabama},\ref{AstroCrete}} \orcid{0000-0001-9200-4006} 
Riccardo Middei \inst{\ref{SSDC_Rome},\ref{INAF_Rome_Obs}} \orcid{0000-0001-9815-9092} 
Dawoon E. Kim \inst{\ref{INAF_Rome},\ref{UniRome1},\ref{UniRome2}} \orcid{0000-0001-5717-3736} 
Fabrizio Tavecchio \inst{\ref{INAF_Merate}} \orcid{0000-0003-0256-0995} 
Alan P. Marscher \inst{\ref{BostonUni}} \orcid{0000-0001-7396-3332} 
Herman L. Marshall \inst{\ref{MIT}} \orcid{0000-0002-6492-1293} 
Steven R. Ehlert \inst{\ref{NASA_Alabama}} \orcid{0000-0003-4420-2838} 
Laura Di Gesu \inst{\ref{Rome_ASI}} \orcid{0000-0002-5614-5028} 
Svetlana G. Jorstad \inst{\ref{BostonUni},\ref{StPetersburg}} \orcid{0000-0001-6158-1708} 
Iv\'{a}n Agudo \inst{\ref{InstAstro_Granada}} \orcid{0000-0002-3777-6182} 
Grzegorz M. Madejski \inst{\ref{Stanford_slac}} 
Roger W. Romani \inst{\ref{Stanford}} \orcid{0000-0001-6711-3286} 
Manel Errando \inst{\ref{StLouis}} \orcid{0000-0002-1853-863X} 
Elina Lindfors \inst{\ref{UTU},\ref{FINCA}} \orcid{0000-0002-9155-6199} 
Kari Nilsson \inst{\ref{FINCA}} \orcid{0000-0002-1445-8683} 
Ella Toppari \inst{\ref{UTU},\ref{FINCA}} 
Stephen B. Potter \inst{\ref{SA_Obs},\ref{SA_UniJ}} 
Ryo Imazawa \inst{\ref{Jap_UniHiroshima}} 
Mahito Sasada \inst{\ref{Jap_TokyoInstTech}} 
Yasushi Fukazawa \inst{\ref{Jap_UniHiroshima},\ref{Jap_ASCHiroshima},\ref{Jap_CoreUHiroshima}} 
Koji S. Kawabata \inst{\ref{Jap_UniHiroshima},\ref{Jap_ASCHiroshima},\ref{Jap_CoreUHiroshima}} 
Makoto Uemura \inst{\ref{Jap_UniHiroshima},\ref{Jap_ASCHiroshima},\ref{Jap_CoreUHiroshima}} 
Tsunefumi Mizuno \inst{\ref{Jap_UniHiroshima}} \orcid{0000-0001-7263-0296} 
Tatsuya Nakaoka \inst{\ref{Jap_ASCHiroshima}} 
Hiroshi Akitaya \inst{\ref{Jap_PERCNarashino}} 
Callum McCall \inst{\ref{LJMU}} 
Helen E. Jermak \inst{\ref{LJMU}} 
Iain A. Steele \inst{\ref{LJMU}} 
Ioannis Myserlis \inst{\ref{IRAM},\ref{MPI_Bonn}} \orcid{0000-0003-3025-9497} 
Mark Gurwell \inst{\ref{Harvard_Smithsonian}} \orcid{0000-0003-0685-3621} 
Garrett K. Keating \inst{\ref{Harvard_Smithsonian}} \orcid{0000-0002-3490-146X} 
Ramprasad Rao \inst{\ref{Harvard_Smithsonian}} 
Sincheol Kang \inst{\ref{Korea_ASSI}} \orcid{0000-0002-0112-4836} 
Sang-Sung Lee \inst{\ref{Korea_ASSI},\ref{Korea_UniSciTech}} \orcid{0000-0002-6269-594X} 
Sang-Hyun Kim \inst{\ref{Korea_ASSI},\ref{Korea_UniSciTech}} \orcid{0000-0001-7556-8504} 
Whee Yeon Cheong \inst{\ref{Korea_ASSI},\ref{Korea_UniSciTech}} \orcid{0009-0002-1871-5824} 
Hyeon-Woo Jeong \inst{\ref{Korea_ASSI},\ref{Korea_UniSciTech}} \orcid{0009-0005-7629-8450} 
Emmanouil Angelakis \inst{\ref{Greece_UniAthens}} \orcid{0000-0001-7327-5441} 
Alexander Kraus \inst{\ref{MPI_Bonn}} \orcid{0000-0002-4184-9372} 
Francisco Jos\'e Aceituno \inst{\ref{InstAstro_Granada}} 
Giacomo Bonnoli \inst{\ref{INAF_Merate},\ref{InstAstro_Granada}} \orcid{0000-0003-2464-9077} 
V\'{i}ctor Casanova \inst{\ref{InstAstro_Granada}} 
Juan Escudero \inst{\ref{InstAstro_Granada}} \orcid{0000-0002-4131-655X} 
Beatriz Ag\'{i}s-Gonz\'{a}lez \inst{\ref{InstAstro_Granada},\ref{AstroCrete}}  
C\'{e}sar Husillos \inst{\ref{geological_madrid},\ref{InstAstro_Granada}} \orcid{0000-0001-8286-5443} 
Daniel Morcuende \inst{\ref{InstAstro_Granada}} 
Jorge Otero-Santos \inst{\ref{InstAstro_Granada}} 
Alfredo Sota \inst{\ref{InstAstro_Granada}} \orcid{0000-0002-9404-6952} 
Rumen Bachev \inst{\ref{Astro_Sofia}} 
Lucio Angelo Antonelli \inst{\ref{INAF_Rome_Obs},\ref{SSDC_Rome}} \orcid{0000-0002-5037-9034} 
Matteo Bachetti \inst{\ref{INAF_Selargius}} \orcid{0000-0002-4576-9337} 
Luca Baldini \inst{\ref{INFN_Pisa},\ref{UniPisa}} \orcid{0000-0002-9785-7726} 
Wayne H. Baumgartner \inst{\ref{NASA_Alabama}} \orcid{0000-0002-5106-0463} 
Ronaldo Bellazzini \inst{\ref{INFN_Pisa}} \orcid{0000-0002-2469-7063} 
Stefano	Bianchi \inst{\ref{UniRome3}} \orcid{0000-0002-4622-4240} 
Stephen	D. Bongiorno \inst{\ref{NASA_Alabama}} \orcid{0000-0002-0901-2097} 
Raffaella Bonino \inst{\ref{INFN_Turin},\ref{UniTurin}} \orcid{0000-0002-4264-1215} 
Alessandro Brez \inst{\ref{INFN_Pisa}} \orcid{0000-0002-9460-1821} 
Niccol\`{o} Bucciantini \inst{\ref{INAF_Florence_Obs},\ref{UniFlorence},\ref{INAF_Florence}} \orcid{0000-0002-8848-1392} 
Fiamma Capitanio \inst{\ref{INAF_Rome}} \orcid{0000-0002-6384-3027} 
Simone Castellano \inst{\ref{INFN_Pisa}} \orcid{0000-0003-1111-4292} 
Elisabetta Cavazzuti \inst{\ref{Rome_ASI}} \orcid{0000-0001-7150-9638} 
Chien-Ting Chen \inst{\ref{USRA_Alabama}} \orcid{0000-0002-4945-5079} 
Stefano Ciprini \inst{\ref{INFN_Rome},\ref{SSDC_Rome}} \orcid{0000-0002-0712-2479} 
Enrico Costa \inst{\ref{INAF_Rome}} \orcid{0000-0003-4925-8523} 
Alessandra De Rosa \inst{\ref{INAF_Rome}} \orcid{0000-0001-5668-6863} 
Ettore Del Monte \inst{\ref{INAF_Rome}} \orcid{0000-0002-3013-6334} 
Niccol\`{o} Di Lalla \inst{\ref{Stanford}} \orcid{0000-0002-7574-1298} 
Alessandro Di Marco	\inst{\ref{INAF_Rome}} \orcid{0000-0003-0331-3259} 
Immacolata Donnarumma \inst{\ref{Rome_ASI}} \orcid{0000-0002-4700-4549} 
Victor Doroshenko \inst{\ref{Tubingen}} \orcid{0000-0001-8162-1105} 
Michal Dov\v{c}iak \inst{\ref{Astro_Prague}} \orcid{0000-0003-0079-1239} 
Teruaki Enoto \inst{\ref{RIKEN}} \orcid{0000-0003-1244-3100} 
Yuri Evangelista \inst{\ref{INAF_Rome}} \orcid{0000-0001-6096-6710} 
Sergio Fabiani \inst{\ref{INAF_Rome}} \orcid{0000-0003-1533-0283} 
Riccardo Ferrazzoli \inst{\ref{INAF_Rome}} \orcid{0000-0003-1074-8605} 
Javier A. Garcia \inst{\ref{NASA_Goddard}} \orcid{0000-0003-3828-2448} 
Shuichi Gunji \inst{\ref{Jap_Yamagata}} \orcid{0000-0002-5881-2445} 
Kiyoshi Hayashida \inst{\ref{UniOsaka}} 
Jeremy Heyl \inst{\ref{UBC}} \orcid{0000-0001-9739-367X} 
Wataru Iwakiri \inst{\ref{Jap_Chiba}} \orcid{0000-0002-0207-9010} 
Philip Kaaret \inst{\ref{NASA_Alabama}} \orcid{0000-0002-3638-0637} 
Vladimir Karas \inst{\ref{Astro_Prague}} \orcid{0000-0002-5760-0459} 
Fabian Kislat \inst{\ref{UNH}} \orcid{0000-0001-7477-0380} 
Takao Kitaguchi \inst{\ref{RIKEN}} 
Jeffery	J. Kolodziejczak \inst{\ref{NASA_Alabama}} \orcid{0000-0002-0110-6136} 
Henric Krawczynski \inst{\ref{StLouis}} \orcid{0000-0002-1084-6507} 
Fabio La Monaca \inst{\ref{INAF_Rome},\ref{UniRome2},\ref{UniRome1}} \orcid{0000-0001-8916-4156} 
Luca Latronico \inst{\ref{INFN_Turin}} \orcid{0000-0002-0984-1856} 
Simone Maldera \inst{\ref{INFN_Turin}} \orcid{0000-0002-0698-4421} 
Alberto Manfreda \inst{\ref{INFN_Naples}} \orcid{0000-0002-0998-4953} 
Fr\'{e}d\'{e}ric Marin \inst{\ref{Strasbourg}} \orcid{0000-0003-4952-0835} 
Andrea Marinucci \inst{\ref{Rome_ASI}} \orcid{0000-0002-2055-4946} 
Francesco Massaro \inst{\ref{INFN_Turin},\ref{UniTurin}} \orcid{0000-0002-1704-9850} 
Giorgio Matt \inst{\ref{UniRome3}} \orcid{0000-0002-2152-0916} 
Ikuyuki Mitsuishi \inst{\ref{Nagoya}} 
Fabio Muleri \inst{\ref{INAF_Rome}} \orcid{0000-0003-3331-3794} 
Michela Negro \inst{\ref{LSU}} \orcid{0000-0002-6548-5622} 
C.-Y. Ng \inst{\ref{HongKong}} \orcid{0000-0002-5847-2612} 
Stephen L. O'Dell \inst{\ref{NASA_Alabama}} \orcid{0000-0002-1868-8056} 
Nicola Omodei \inst{\ref{Stanford}} \orcid{0000-0002-5448-7577} 
Chiara Oppedisano \inst{\ref{INFN_Turin}} \orcid{0000-0001-6194-4601} 
Alessandro Papitto \inst{\ref{INAF_Rome_Obs}} \orcid{0000-0001-6289-7413} 
George G. Pavlov \inst{\ref{PennState}} \orcid{0000-0002-7481-5259} 
Abel Lawrence Peirson \inst{\ref{Stanford}} \orcid{0000-0001-6292-1911} 
Matteo Perri \inst{\ref{SSDC_Rome},\ref{INAF_Rome_Obs}} \orcid{0000-0003-3613-4409} 
Melissa Pesce-Rollins \inst{\ref{INFN_Pisa}} \orcid{0000-0003-1790-8018} 
Pierre-Olivier Petrucci \inst{\ref{Grenoble}} \orcid{0000-0001-6061-3480} 
Maura Pilia \inst{\ref{INAF_Selargius}} \orcid{0000-0001-7397-8091} 
Andrea Possenti \inst{\ref{INAF_Selargius}} \orcid{0000-0001-5902-3731} 
Juri Poutanen \inst{\ref{UTU}} \orcid{0000-0002-0983-0049} 
Simonetta Puccetti \inst{\ref{SSDC_Rome}} \orcid{0000-0002-2734-7835} 
Brian D. Ramsey \inst{\ref{NASA_Alabama}} \orcid{0000-0003-1548-1524} 
John Rankin \inst{\ref{INAF_Rome}} \orcid{0000-0002-9774-0560} 
Ajay Ratheesh \inst{\ref{INAF_Rome}} \orcid{0000-0003-0411-4243} 
Oliver J. Roberts \inst{\ref{NASA_Alabama}} \orcid{0000-0002-7150-9061} 
Carmelo Sgr\`{o} \inst{\ref{INFN_Pisa}} \orcid{0000-0001-5676-6214} 
Patrick Slane \inst{\ref{Harvard_Smithsonian}} \orcid{0000-0002-6986-6756} 
Paolo Soffitta \inst{\ref{INAF_Rome}} \orcid{0000-0002-7781-4104} 
Gloria Spandre \inst{\ref{INFN_Pisa}} \orcid{0000-0003-0802-3453} 
Douglas A. Swartz \inst{\ref{NASA_Alabama}} \orcid{0000-0002-2954-4461} 
Toru Tamagawa \inst{\ref{RIKEN}} \orcid{0000-0002-8801-6263} 
Roberto Taverna \inst{\ref{UniPadova}} \orcid{0000-0002-1768-618X} 
Yuzuru Tawara \inst{\ref{Nagoya}} 
Allyn F. Tennant \inst{\ref{NASA_Alabama}} \orcid{0000-0002-9443-6774} 
Nicholas E. Thomas \inst{\ref{NASA_Alabama}} \orcid{0000-0003-0411-4606} 
Francesco Tombesi \inst{\ref{UniRome2},\ref{INFN_Rome},\ref{UniMaryland}} \orcid{0000-0002-6562-8654} 
Alessio Trois \inst{\ref{INAF_Selargius}} \orcid{0000-0002-3180-6002} 
Sergey S. Tsygankov \inst{\ref{UTU}} \orcid{0000-0002-9679-0793} 
Roberto Turolla \inst{\ref{UniPadova},\ref{MSSL}} \orcid{0000-0003-3977-8760} 
Jacco Vink \inst{\ref{Amsterdam}} \orcid{0000-0002-4708-4219} 
Martin C. Weisskopf \inst{\ref{NASA_Alabama}} \orcid{0000-0002-5270-4240} 
Kinwah Wu \inst{\ref{MSSL}} \orcid{0000-0002-7568-8765} 
Fei Xie \inst{\ref{China_Guangxi},\ref{INAF_Rome}} \orcid{0000-0002-0105-5826} 
Silvia Zane \inst{\ref{MSSL}} \orcid{0000-0001-5326-880X} 
}

\institute{
Department of Physics and Astronomy, University of Turku, FI-20014, Finland \label{UTU}
\and
Finnish Centre for Astronomy with ESO (FINCA), Quantum, Vesilinnantie 5, FI-20014 University of Turku, Finland \label{FINCA}
\and
Aalto University Mets\"ahovi Radio Observatory, Mets\"ahovintie 114, FI-02540 Kylm\"al\"a, Finland \label{MRO}
\and
NASA Marshall Space Flight Center, Huntsville, AL 35812, USA \label{NASA_Alabama}
\and
Institute of Astrophysics, Foundation for Research and Technology-Hellas, GR-70013 Heraklion, Greece \label{AstroCrete} 
\and
INAF Istituto di Astrofisica e Planetologia Spaziali, Via del Fosso del Cavaliere 100, 00133 Roma, Italy \label{INAF_Rome}
\and
Dipartimento di Fisica, Universit\`{a} degli Studi di Roma "La Sapienza", Piazzale Aldo Moro 5, 00185 Roma, Italy\label{UniRome1} 
\and
Dipartimento di Fisica, Universit\`{a} degli Studi di Roma "Tor Vergata", Via della Ricerca Scientifica 1, 00133 Roma, Italy \label{UniRome2} 
\and
Space Science Data Center, Agenzia Spaziale Italiana, Via del Politecnico snc, 00133 Roma, Italy \label{SSDC_Rome} 
\and
INAF Osservatorio Astronomico di Roma, Via Frascati 33, 00078 Monte Porzio Catone (RM), Italy \label{INAF_Rome_Obs} 
\and
INAF Osservatorio Astronomico di Cagliari, Via della Scienza 5, 09047 Selargius (CA), Italy \label{INAF_Selargius}
\and
INAF Osservatorio Astronomico di Brera, Via E. Bianchi 46, 23807 Merate (LC), Italy \label{INAF_Merate}
\and
Institute for Astrophysical Research, Boston University, 725 Commonwealth Avenue, Boston, MA 02215, USA \label{BostonUni}
\and
MIT Kavli Institute for Astrophysics and Space Research, Massachusetts Institute of Technology, 77 Massachusetts Avenue, Cambridge, MA 02139, USA \label{MIT}
\and
South African Astronomical Observatory, PO Box 9, Observatory, 7935, Cape Town, South Africa \label{SA_Obs}
\and
Department of Physics, University of Johannesburg, PO Box 524, Auckland Park 2006, South Africa \label{SA_UniJ}
\and
Department of Physics, Graduate School of Advanced Science and Engineering, Hiroshima University Kagamiyama, 1-3-1 Higashi-Hiroshima, Hiroshima 739-8526, Japan \label{Jap_UniHiroshima}
\and
Hiroshima Astrophysical Science Center, Hiroshima University 1-3-1 Kagamiyama, Higashi-Hiroshima, Hiroshima 739-8526, Japan\label{Jap_ASCHiroshima}
\and
Department of Physics, Tokyo Institute of Technology, 2-12-1 Ookayama, Meguro-ku, Tokyo 152-8551, Japan \label{Jap_TokyoInstTech}
\and
Core Research for Energetic Universe (Core-U), Hiroshima University, 1-3-1 Kagamiyama, Higashi-Hiroshima, Hiroshima 739-8526, Japan \label{Jap_CoreUHiroshima}
\and
Planetary Exploration Research Center, Chiba Institute of Technology 2-17-1 Tsudanuma, Narashino, Chiba 275-0016, Japan \label{Jap_PERCNarashino}
\and
Institut de Radioastronomie Millim\'{e}trique, Avenida Divina Pastora, 7, Local 20, E–18012 Granada, Spain \label{IRAM}
\and
Max-Planck-Institut f\"{u}r Radioastronomie, Auf dem H\"{u}gel 69, D-53121 Bonn, Germany \label{MPI_Bonn}
\and
Center for Astrophysics, Harvard \& Smithsonian, 60 Garden Street, Cambridge, MA 02138 USA \label{Harvard_Smithsonian}
\and
Korea Astronomy and Space Science Institute, 776 Daedeok-daero, Yuseong-gu, Daejeon 34055, Korea \label{Korea_ASSI}
\and
University of Science and Technology, Korea, 217 Gajeong-ro, Yuseong-gu, Daejeon 34113, Korea \label{Korea_UniSciTech}
\and
Section of Astrophysics, Astronomy \& Mechanics, Department of Physics, National and Kapodistrian University of Athens, Panepistimiopolis Zografos 15784, Greece \label{Greece_UniAthens}
\and
Instituto de Astrof\'{i}sica de Andaluc\'{i}a, IAA-CSIC, Glorieta de la Astronom\'{i}a s/n, 18008 Granada, Spain \label{InstAstro_Granada}
\and
Istituto Nazionale di Fisica Nucleare, Sezione di Pisa, Largo B. Pontecorvo 3, 56127 Pisa, Italy \label{INFN_Pisa}
\and
Dipartimento di Fisica, Universit\`{a} di Pisa, Largo B. Pontecorvo 3, 56127 Pisa, Italy \label{UniPisa}
\and
Dipartimento di Matematica e Fisica, Universit\`{a} degli Studi Roma Tre, Via della Vasca Navale 84, 00146 Roma, Italy \label{UniRome3}
\and
Istituto Nazionale di Fisica Nucleare, Sezione di Torino, Via Pietro Giuria 1, 10125 Torino, Italy \label{INFN_Turin}
\and
Dipartimento di Fisica, Universit\`{a} degli Studi di Torino, Via Pietro Giuria 1, 10125 Torino, Italy \label{UniTurin}
\and
INAF Osservatorio Astrofisico di Arcetri, Largo Enrico Fermi 5, 50125 Firenze, Italy \label{INAF_Florence_Obs}
\and
Dipartimento di Fisica e Astronomia, Universit\`{a} degli Studi di Firenze, Via Sansone 1, 50019 Sesto Fiorentino (FI), Italy \label{UniFlorence}
\and
Istituto Nazionale di Fisica Nucleare, Sezione di Firenze, Via Sansone 1, 50019 Sesto Fiorentino (FI), Italy \label{INAF_Florence}
\and
ASI - Agenzia Spaziale Italiana, Via del Politecnico snc, 00133 Roma, Italy \label{Rome_ASI}
\and
Science and Technology Institute, Universities Space Research Association, Huntsville, AL 35805, USA \label{USRA_Alabama}
\and
Istituto Nazionale di Fisica Nucleare, Sezione di Roma "Tor Vergata", Via della Ricerca Scientifica 1, 00133 Roma, Italy \label{INFN_Rome}
\and
Department of Physics and Kavli Institute for Particle Astrophysics and Cosmology, Stanford University, Stanford, California 94305, USA \label{Stanford}
\and
Kavli Institute for Particle Astrophysics and Cosmology, Stanford University, and SLAC 2575 Sand Hill Road, Menlo Park, CA 94025, USA \label{Stanford_slac}
\and
Institut f\"{u}r Astronomie und Astrophysik, Universit\"{a}t T\"{u}bingen, Sand 1, 72076 T\"{u}bingen, Germany \label{Tubingen}
\and
Astronomical Institute of the Czech Academy of Sciences, Bo\v{c}n\'{i} II 1401/1, 14100 Praha 4, Czech Republic \label{Astro_Prague}
\and
RIKEN Cluster for Pioneering Research, 2-1 Hirosawa, Wako, Saitama 351-0198, Japan \label{RIKEN}
\and
NASA Goddard Space Flight Center, Greenbelt, MD 20771, USA \label{NASA_Goddard}
\and
Yamagata University, 1-4-12 Kojirakawa-machi, Yamagata-shi 990-8560, Japan \label{Jap_Yamagata}
\and
Osaka University, 1-1 Yamadaoka, Suita, Osaka 565-0871, Japan \label{UniOsaka}
\and
University of British Columbia, Vancouver, BC V6T 1Z4, Canada \label{UBC}
\and
International Center for Hadron Astrophysics, Chiba University, Chiba 263-8522, Japan \label{Jap_Chiba}
\and
St. Petersburg State University, 7/9, Universitetskaya nab., 199034 St. Petersburg, Russia \label{StPetersburg}
\and
Department of Physics and Astronomy and Space Science Center, University of New Hampshire, Durham, NH 03824, USA \label{UNH}
\and
Physics Department and McDonnell Center for the Space Sciences, Washington University in St. Louis, St. Louis, MO 63130, USA \label{StLouis}
\and
Istituto Nazionale di Fisica Nucleare, Sezione di Napoli, Strada Comunale Cinthia, 80126 Napoli, Italy \label{INFN_Naples}
\and
Universit\'{e} de Strasbourg, CNRS, Observatoire Astronomique de Strasbourg, UMR 7550, 67000 Strasbourg, France \label{Strasbourg}
\and
Graduate School of Science, Division of Particle and Astrophysical Science, Nagoya University, Furo-cho, Chikusa-ku, Nagoya, Aichi 464-8602, Japan \label{Nagoya}
\and
Department of Physics and Astronomy, Louisiana State University, Baton Rouge, LA 70803, USA \label{LSU}
\and
Department of Physics, The University of Hong Kong, Pokfulam, Hong Kong \label{HongKong}
\and
Department of Astronomy and Astrophysics, Pennsylvania State University, University Park, PA 16802, USA \label{PennState}
\and
Universit\'{e} Grenoble Alpes, CNRS, IPAG, 38000 Grenoble, France \label{Grenoble}
\and
Dipartimento di Fisica e Astronomia, Universit\`{a} degli Studi di Padova, Via Marzolo 8, 35131 Padova, Italy \label{UniPadova}
\and
Department of Astronomy, University of Maryland, College Park, Maryland 20742, USA \label{UniMaryland}
\and
Mullard Space Science Laboratory, University College London, Holmbury St Mary, Dorking, Surrey RH5 6NT, UK \label{MSSL}
\and
Anton Pannekoek Institute for Astronomy \& GRAPPA, University of Amsterdam, Science Park 904, 1098 XH Amsterdam, The Netherlands \label{Amsterdam}
\and
Guangxi Key Laboratory for Relativistic Astrophysics, School of Physical Science and Technology, Guangxi University, Nanning 530004, China \label{China_Guangxi}
\and
Institute of Astronomy and NAO, Bulgarian Academy of Sciences, 1784 Sofia, Bulgaria \label{Astro_Sofia}
\and
Astrophysics Research Institute, Liverpool John Moores University, Liverpool Science Park IC2, 146 Brownlow Hill, UK \label{LJMU}
\and
Geological and Mining Institute of Spain (IGME-CSIC), Calle R\'{i}os Rosas 23, E-28003, Madrid, Spain \label{geological_madrid}
}


\date{Received January 05, 2024; accepted May 30, 2024}

 
\abstract{
We report the X-ray polarization properties of the high-synchrotron-peaked (HSP) blazar PKS~2155$-$304 based on observations with the Imaging X-ray Polarimetry Explorer (IXPE). We observed the source between Oct 27 and Nov 7, 2023. We also conducted an extensive contemporaneous multiwavelength (MW) campaign. We find that during the first half ($T_1$) of the IXPE pointing, the source exhibited the highest X-ray polarization degree detected for an HSP blazar thus far, (30.7$\pm$2.0)\%, which dropped to (15.3$\pm$2.1)\% during the second half ($T_2$). The X-ray polarization angle remained stable during the IXPE pointing at 129.4\degree$\pm$1.8\degree\, and 125.4\degree$\pm$3.9\degree\, during $T_1$ and $T_2$, respectively. Meanwhile, the optical polarization degree remained stable during the IXPE pointing, with average host-galaxy-corrected values of (4.3$\pm$0.7)\% and (3.8$\pm$0.9)\% during the $T_1$ and $T_2$, respectively. During the IXPE pointing, the optical polarization angle changed achromatically from $\sim$140\degree\, to $\sim$90\degree\, and back to $\sim$130\degree. Despite several attempts, we only detected (99.7\% conf.) the radio polarization once (during $T_2$, at 225.5~GHz): with degree (1.7$\pm$0.4)\% and angle 112.5\degree$\pm$5.5\degree. The direction of the broad pc-scale jet is rather ambiguous and has been found to point to the east and south at different epochs; however, on larger scales (>~1.5~pc) the jet points toward the southeast ($\sim$135\degree), similar to all of the MW polarization angles. Moreover, the X-ray to optical polarization degree ratios of $\sim$7 and $\sim$4 during $T_1$ and $T_2$, respectively, are similar to previous IXPE results for several HSP blazars. These findings, combined with the lack of correlation of temporal variability between the MW polarization properties, agree with an energy-stratified shock-acceleration scenario in HSP blazars.
}

\keywords{BL Lacertae objects: HSP --Galaxies: jets --Polarization --Relativistic processes --Magnetic fields} 

\titlerunning{IXPE observation of PKS2155$-$304 reveals the most polarized blazar}
\authorrunning{Kouch et al.}

\maketitle
%

\section{Introduction} \label{sec:intro}

The supermassive black holes in the centers of active galactic nuclei (AGN)  sometimes (in 5-10\% of AGN) launch highly relativistic plasma jets that emit extremely luminous non-thermal radiation. These jets serve as laboratories for studying acceleration, cooling, and interactions of particles in some of the most energetically extreme environments in the Universe, involving ultra-relativistic electrons with Lorentz factors $\gtrsim$$10^6$. A particularly prominent subclass of AGN for studying such phenomena is blazars, whose plasma jets are well aligned with our line of sight (e.g., \citealt{Blandford2019,Hovatta2019}). Blazars appear exceptionally luminous due to relativistic beaming and exhibit a Doppler-boosted double-hump spectral energy distribution (SED) that stretches from the radio to $\gamma$-ray bands. The first hump is generally attributed to synchrotron emission of charged particles accelerating in the magnetized jet. The origin of the second hump is not yet fully understood, although Compton up-scattering of lower-energy photons is likely to contribute significantly to it (e.g., \citealt{Paliya2018}). Blazars are commonly categorized based on their synchrotron peak frequency into low, intermediate, and high-synchrotron peaked sources: LSP ($\nu_{\mathrm{synch}}$<$10^{14}$~Hz), ISP ($10^{14}$<$\nu_{\mathrm{synch}}$<$10^{15}$~Hz), and HSP ($\nu_{\mathrm{synch}}$>$10^{15}$~Hz). As such, HSP blazars have synchrotron peaks that correspond to the UV/X-ray photon energy range ($\gtrsim$0.01~keV).

Multiwavelength polarization measurements of blazar emission is a vital research tool to distinguish among the predictions of different particle acceleration and emission models. The polarization degree ($\Pi$=$\sqrt{\mathrm{Q}^2+\mathrm{U}^2+\mathrm{V}^2}/\mathrm{I}$, where Q, U, V, and I are the Stokes parameters), which measures the fraction of polarized radiation, reveals the level of order of the magnetic field lines in the emission region. The polarization angle
($\Psi$=$\frac{1}{2}\arctan\left[ \frac{\mathrm{U}}{\mathrm{Q}} \right]$), which refers to the direction of the electric vector of the linearly polarized emission, indicates the orientation of the mean magnetic field in the emission region when the synchrotron-emitting particles are distributed isotropically in the jet frame. Moreover, comparing the magnitude and the temporal variability of the multiwavelength polarization measurements ($\Pi$ and $\Psi$) can determine whether the emission at different wave bands is co-spatial. Radio and optical polarization measurements in the past have been used to constrain emission models in LSP and ISP blazars (e.g., \citealt{Marscher2008_nature_MW_EVPA_rot}). However, in the case of HSP sources, whose peak luminosity lies in the UV/X-ray domain, polarization measurements at higher photon energies are more advantageous as research tools. 

Since the first observation of Cen~A in early 2022 \citep{Ehlert2022CenA}, the Imaging X-ray Polarimetry Explorer (IXPE, \citealp{Weisskopf2022_ixpe_technical}) has opened a new window to the extragalactic Universe by obtaining polarization measurements in the medium-energy X-ray range (2\textendash8~keV). For HSPs, this range means that IXPE probes their synchrotron dominated emissions. The first blazar observations, of the HSPs Mrk~501 and Mrk~421, found that the synchrotron X-ray polarization degree ($\Pi_\mathrm{X}$) is a factor of a few times higher than the contemporaneous radio and optical values \citep{Liodakis2022nature,DiGesu2022-Mrk421}. Observations of other HSP sources (e.g., \citealt{Ehlert2023_ixpe_1es0229,Kim2024}) have revealed that the ratio can be up to a factor of seven. The polarization angle $\Psi$ tends to be aligned with the jet axis on the plane of the sky, although large rotations up to 400$\degree$ have been observed at both X-ray and optical wavelengths during the IXPE observations \citep{DiGesu2023,Middei2023-II}. The wavelength dependence of $\Pi$, where $\Pi_\mathrm{X}$>$\Pi_\mathrm{O}$$\gtrsim$$\Pi_\mathrm{R}$, and the lack of correlation between the polarization patterns at different wavelengths, have been interpreted as evidence for an energy-stratified shock-acceleration scenario \citep{Liodakis2022nature, Ehlert2023_ixpe_1es0229}. In this scenario, a compact shock front, possibly swirling down a helical magnetic field in the jet, accelerates particles away from it. The particles lose energy (e.g., via synchrotron emission) as they move farther away into more turbulent regions. As a result, the higher energy emission originates from closer to the shock front where the magnetic field is more ordered, leading to the higher energy emission being more polarized. On the other hand, the lower energy emission originates from magnetically more turbulent regions, leading to them being less polarized. However, other IXPE observations (e.g., 1ES~1959+650;\citealt{Errando2024_1ES1959_ixpe}) have shown comparable polarization degrees in the X-ray and optical bands. This would suggest that a significant turbulent component could be present, even in the X-ray emission region closer to the shock front.

In this paper we present a study of the sixth HSP blazar observed by IXPE, PKS~2155$-$304 ($\alpha$=21:58:52.065, $\delta$=$-$30:13:32.118). It is one of the brightest extragalactic UV and X-ray sources in the southern sky, with a redshift of 0.117 (\citealt{Bowyer1984_redshift}). Since its detection in the keV X-ray band (\citealt{Schwartz1979_X_discovery,Griffiths1979_X_discovery}) where it exhibits extreme variability, PKS~2155$-$304 has been the target of several multiwavelength campaigns from radio to very high-energy (VHE; E>100~GeV) $\gamma$-rays (see, e.g., \citealt{HESS2012_MW_view} and references therein). It has shown intra-night variability in the optical (\citealt{Carini1992_opt_var}), X-ray, and VHE bands, with its spectrum showing a harder-when-brighter behavior (\citealt{Aharonian2005_x_gamma_var}). While a temporal correlation of flux variations in the optical and VHE bands has been suggested, a lack of such correlation between the X-ray and VHE bands challenges the commonly favored one-zone synchrotron self-Compton (SSC) acceleration mechanism used to explain the second SED hump of HSP blazars (\citealt{Aharonian2009_low_state_MW_campaign}). The extremely short variability timescales in the higher-energy bands suggest a much larger Doppler factor (\citealt{Aharonian2007_extreme_Dopfac}) than implied by VLBI observations of the parsec-scale jet (\citealt{Piner2004_VLBI,Piner2008_VLBI_followup}). This may be partly due to the radio and higher-energy emission originating from different parts of the jet. Nevertheless, this issue has prompted the consideration of jet-in-jet models involving magnetic reconnection (\citealt{Giannios2009_jet_in_jet}), spine-sheath models (e.g., \citealt{Ghisellini2005_spine_sheath}), and decelerating flow models (\citealt{Georganopoulos2003_decelerating_flow}).

As observed with very long-baseline interferometry (VLBI) in 2000 at 15~GHz, the projected direction of the jet on PKS~2155$-$304 into the plane of the sky appeared to point due east of the compact core at the smallest scales, $<1.5$~pc (\citealt{Piner2004_VLBI}). However, higher-resolution VLBI images taken in 2009 at 43~GHz revealed that the jet at these smallest scales is oriented southward (\citealt{Piner2010_VLBI_jet_bending}). In the images at both frequencies, the jet points in the southeast direction ($\sim$135\degree) at distances $>1.5$~pc (in projection on the sky) from the compact "core". Due to this ambiguity, we approximate the general direction and opening angle of the jet to encompass the southeastern quadrant between 90\degree\textendash180\degree. Given that the aforementioned VLBI images are not contemporaneous, and that Doppler effects can cause rapid apparent position angle variation of the highly aligned inner jets of blazars, as well as that the jet could have physically changed direction (as many blazars do; see, e.g., \citealt{Lico2020_pc_jet_wobble}), we consider the average jet direction to be better represented by the more stable position angle at the larger scale ($>1.5$~pc). Thus, we estimate the jet position angle to be $\sim$135\degree with an uncertainty of $\pm$30\degree, which is similar to the broad jets found in other HSP sources (\citealt{Weaver2022_jet_PA_list}).

In the optical band, the linear polarization of PKS~2155$-$304 was tracked for around 10 years at Steward Observatory (\citealt{Smith2009_steward_opt_pol_long_term}). The polarization degree $\Pi_\mathrm{O}$ was typically around 2\textendash6\% and the position angle $\Psi_\mathrm{O}$ ranged from 60$\degree$ to 120$\degree$\,; see Appendix \ref{appendix:longterm}.

In \S\ref{sec:xray_pol} we present the observations with IXPE and other X-ray observatories, describe the X-ray data analysis, and report the results. In \S\ref{sec:mwl} we present the data from our contemporaneous multiwavelength campaign. We discuss our findings in \S\ref{sec:disc} and present our conclusions in \S\ref{sec:conc}.

\section{X-ray polarization observations \& analysis}\label{sec:xray_pol}

\subsection{IXPE}\label{sec:xray_pol_ixpe}
IXPE, a joint space mission of NASA and the Italian Space Agency (ASI), comprises three identical X-ray telescopes designed for X-ray polarimetry in the 2\textendash8~keV energy range \citep{Weisskopf2022_ixpe_technical}. For each photon, the detectors generate an electron track from which the Stokes Q, U, and I parameters can be estimated\footnote{Current technology to measure X-ray polarization does not permit measurements of the circular polarization component (Stokes V). Nevertheless, intrinsic circular polarization in blazars should be low  ($\ll$1\%), and has yet to be confidently detected \citep{Liodakis2022circular,Liodakis2023circular}.}. After correcting for instrumental effects (modulation factor, boom motion, etc.), the Stokes parameters are extracted as a function of energy for a given region of interest. Then, the polarization properties (degree and angle) are modeled and estimated using standard X-ray analysis procedures in \texttt{XSPEC}.

IXPE targeted PKS~2155$-$304 from UT 2023-10-27T16:31 to 2023-11-07T00:08. The total exposure time of the observations was 476~ks. The spectra for the Stokes parameters I, Q, and U were derived using a circular region with a radius of 0.95~arcmin centered on the source. An annulus with an inner ring radius of 1.2~arcmin and an outer ring radius of 3.5~arcmin was used to extract the background. The adoption of these regions led us to a total count of around $10^5$ (summing those collected in the three distinct Detector Units). The background contribution is of the order of 2.5\%. Finally, the Stokes I spectra were grouped in order to ensure a signal-to-noise ratio (S/N) of at least 7 in each energy bin, while for the Stokes Q and U spectra we adopted a uniform bin width of 280~eV.

\subsection{\textit{XMM-Newton}}\label{sec:xray_pol_xmm}
On 2023-11-01, \textit{XMM-Newton} (with a sensitivity range of 0.3\textendash10~keV; \citealt{Jansen2001_xmm}) observed PKS~2155$-$304 quasi-simultaneously with IXPE for $\sim$50~ks. Using the standard System Analysis Software (SAS, version xmmsas\_20230412\_1735-21.0.0) and the updated current calibration files, we extracted the event lists of the European Photon Imaging Camera (EPIC; \citealt{Struder2001_xmm_EPIC}). Since the observation was performed in Timing mode, we extracted the source spectrum using a box of 27 pixels centered on the source. The standard SAS commands \textit{rmfgen} and \textit{arfgen} were used to generate the response and auxiliary matrices. The spectrum was subsequently binned so as to avoid oversampling the spectral resolution by a factor greater than 3, and to have at least 30 counts per bin. The obtained spectrum has $2.1 \times 10^6$ counts, with the background being around 4.5\%.

\subsection{\textit{Swift}-XRT}\label{sec:xray_pol_swift}
The \textit{Neil Gehrels Swift} X-Ray Telescope (\textit{Swift}-XRT; sensitive range of 0.2\textendash10~keV; \citealt{Burrows2005_swift}), observed PKS~2155$-$304 daily from one day before to one day after the $\sim10$-day IXPE exposure, and less frequently before then, starting in 2023 September. The information from \textit{Swift} enables us to monitor the evolution of the flux level and the spectral shape of the source. To derive third-level science products, we used the XRT Data Analysis Software with the most recent (2023 July 25) calibration files found in the \textit{Swift}-XRT CALDB. \textit{Swift} observed our source both in window timing (WT) and photon counting (PC) modes. For each observation performed in WT observing mode, we computed the XRT spectrum using the cleaned event files, adopting an extraction region of circular shape with a 47~arcsec radius. The background was then obtained using a blank sky WT observation available in the \textit{Swift} archive. No pile-up issues affect these observations. To mitigate any possible pile-up issues of the PC exposures, the source spectrum was extracted using an annular region with a fixed outer radius of 47~arcsec and a variable inner radius. To mitigate any spectral hardening due to the pile-up, we selected the radius based on Table 2 of \cite{Middei2022}. The background was instead extracted using a circular region (r$\sim$47 arcsec) centered on a black area of the detector. The resulting spectra were subsequently binned, requiring at least 25 counts per bin. The \textit{Swift}-XRT observations presented here are all snapshots of $\sim$1~ks in duration. As a result, each spectrum has a few thousands of counts with a background of the order of $\sim$1\% of the total counts.

We then performed a standard spectral analysis with the \texttt{XSPEC} software \citep{Arnaud1996} by testing each of the resulting spectra using a (tbabs~$\times$~logpar) model. This model accounts for a fixed Galactic column density and a logarithmic parabola commonly used to model the continuum spectrum of HSP objects (see \S\ref{sec:xray_specpol}). The two-month long \textit{Swift}-XRT X-ray light curve of PKS~2155$-$304 is presented in Appendix \ref{appendix:longterm}. According to this light curve, PKS~2155$-$304 was in a typical flux state in the X-ray band during the IXPE pointing.

\subsection{X-ray spectro-polarimetric analysis}\label{sec:xray_specpol}
We derived the X-ray polarimetric properties by performing a spectro-polarimetric analysis using \texttt{XSPEC}. The X-ray spectra of HSP sources are commonly fit with a log-parabola model (e.g., \citealt{Massaro2004,Giommi2021,Middei2022blaz}), hence we modelled the I, Q, and U IXPE spectra as well as the EPIC-pn spectrum using a \texttt{tbabs~$\times$~const$~\times$~logpar$~\times~$polconst} model. The first component, \texttt{tbabs}, accounts for the Galactic hydrogen column density, which was set to a fixed value of $N_{\rm H}$=1.29$\times$10$^{20}$ cm$^{-2}$ (\citealt{HI4PI2016}). The following component, \texttt{const}, is a multiplicative constant needed to take into account the inter-calibration among the different detector units (DUs) and the EPIC-pn camera spectrum. The component \texttt{logpar} is used when fitting the curved photon-flux continuum with a logarithmic-parabolic equation: $\Phi(E) = K~(E/E_\mathrm{p})^{-\alpha-\beta\log(E/E_\mathrm{p})}$, where $E_\mathrm{p}$ is a scaling factor referred to as the pivot energy, $K$ is a normalization factor, $\alpha$ is the slope at the pivot energy, and $\beta$ is the curvature term (e.g., \citealt{Massaro2004}). Finally, the last model component, \texttt{polconst}, accounts for the polarimetric properties of the Q and U spectra, thus returning $\rm \Pi_X$ and $\rm \Psi_X$ fits under the assumption that both of these quantities remained constant over the IXPE operating energy range.

During the fitting procedure, we initially computed both $\alpha$ and $\beta$ of \texttt{logpar} while assuming that they do not vary between the IXPE and \textit{XMM-Newton} exposures. Similarly, we fitted the normalization of the continuum ($K$) and the different cross-calibration constants (k$_{\rm DU2}$, k$_{\rm DU3}$, and k$_{\rm pn}$). Additionally, $E_\mathrm{p}$ was set to 3~keV. However, this approach does not return a good fit, as a preliminary analysis revealed the IXPE and \textit{XMM-Newton} slopes ($\alpha$) to differ by $\sim$0.1. This can be either due to spectral calibration issues, intrinsic changes of the spectral shape of PKS~2155$-$304 (e.g., \citealt{Middei2022blaz}), or the lack of truly simultaneous data. To account for this, we fit $\alpha$ separately for the IXPE and \textit{XMM-Newton} data.

These steps yield a logarithmic parabola best-fit with $\chi^2$/d.o.f.=596/538, i.e., the chance probability for the data to have been drawn from the model is 0.04. Despite several studies discussing the importance of modelling the spectra of HSP sources with a logarithmic parabola model (e.g., \citealt{Massaro2004}), we additionally checked the goodness-of-fit of our IXPE-\textit{XMM-Newton} data with an absorbed power-law model. We did this by replacing the logarithmic parabola component in our model with a new free parameter (\texttt{tbabs}) plus a power-law. For the last component we calculated the photon index ($\Gamma$) of the IXPE and \textit{XMM-Newton} datasets separately. The $\chi^2$ goodness-of-fit test for the absorbed power-law model results in $\chi^2$/d.o.f.=626/538 (with a 0.005 chance probability of the data having been drawn from the model), which is a worse fit than the logarithmic parabola model.

\begin{figure*}
    \centering
    \includegraphics[width=\textwidth]{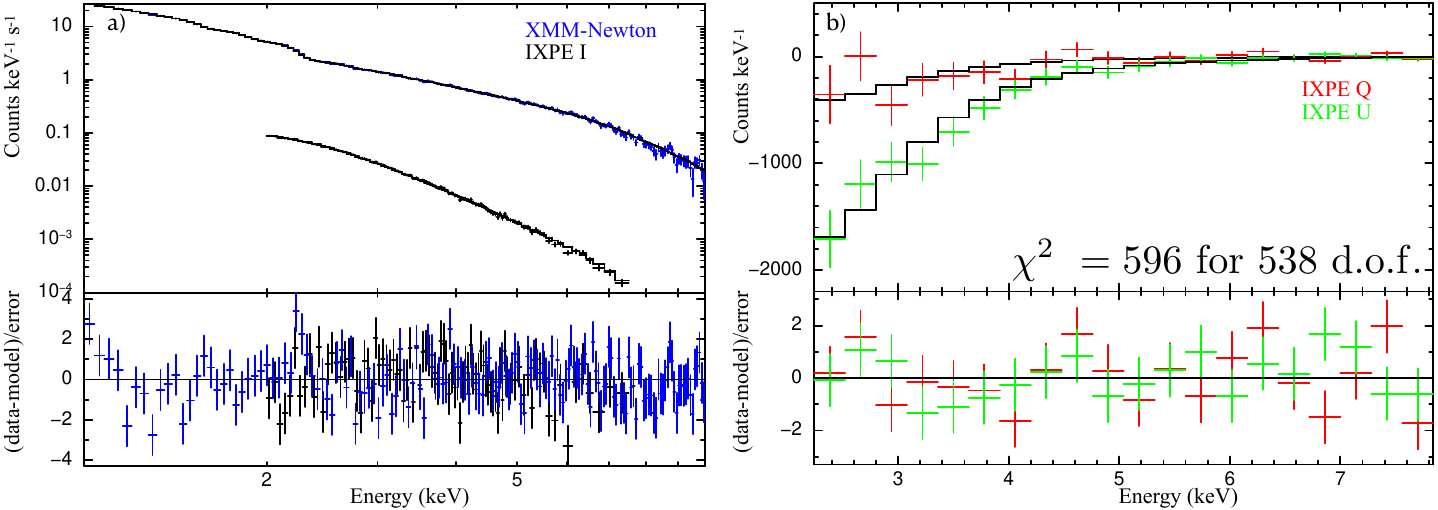}
    \caption{\textit{Top panels:} Best fits to the IXPE and \textit{XMM-Newton} spectra; \textit{bottom panels:} The corresponding residuals. \textit{Left:} The fit of the model \texttt{tbabs~$\times$~const~$\times$~polconst~$\times$~logpar} to the Stokes I spectra; \textit{right:} The best-fit Stokes Q and U spectra.}
    \label{xspecpol}
\end{figure*}

In Figure~\ref{xspecpol}, panel (a), we show how the logarithmic parabola model fits the data and the corresponding residuals for the I Stokes parameters. The best-fit Q and U Stokes parameters are shown in panel (b) of Figure~\ref{xspecpol}. Table~\ref{bestfitTable} gives the inferred parameters and their uncertainties.

\begin{table}
\setlength{\tabcolsep}{1.9pt}
\centering
\caption{Spectro-polarimetric best-fit parameters of PKS~2155$-$304.\label{bestfitTable}}
	\begin{tabular}{l l r l}
	\hline\hline
	Component & Parameter & Time-averaged & Units  \\
	\hline\hline
    \texttt{polconst} & $\Pi_\mathrm{X}$ & 23.3$\pm$1.5 & (\%) \\
     &  $\Psi_\mathrm{X}$ & 128$\pm1.8$ & (\degree) \\
     \hline
    \texttt{tbabs} & n$_{\rm H}$$\dagger$ & 1.29 & 10$^{20}$ cm$^{-2}$ \\
    \hline
    \texttt{log-par} & $\alpha$ & 2.71$\pm0.02$ & \\
     & $\alpha_{\rm pn}$ & 2.569$\pm0.08$ & \\
    & $E_\mathrm{p}\dagger$ & 3 & keV \\
    & $\beta$ & 0.078$\pm$0.012 &  \\
    & $K$ &$3.2\pm0.1$ &  \\
    \hline
    \texttt{const} & k$_{\rm DU2}$ & 0.96$\pm$0.01 & \\
     & k$_{\rm DU3}$ & 0.89$\pm$0.01 &  \\
     & k$_{\rm pn}$ & 0.96$\pm$0.01 & \\
    \hline
    $F_{\rm 2-8~keV}$&&2.63$\pm$0.01 & $10^{-11}$ erg cm$^{-2}$ s$^{-1}$ \\
    \hline
	\end{tabular}
    \tablefoot{The results refer to the joint fit to the IXPE and \textit{XMM-Newton} data. Fluxes are given in units of $10^{-11}$ erg cm$^{-2}$ s$^{-1}$, while the logarithmic parabola has a normalization of $10^{-2}$. All errors are given at 68\% confidence for one parameter of interest (i.e., $\Delta \chi^2$=1). The symbol $\dagger$ indicates those parameters that were kept frozen during the model fitting.}
\end{table}

\begin{figure}
    \centering
    \includegraphics[width=0.49\textwidth]{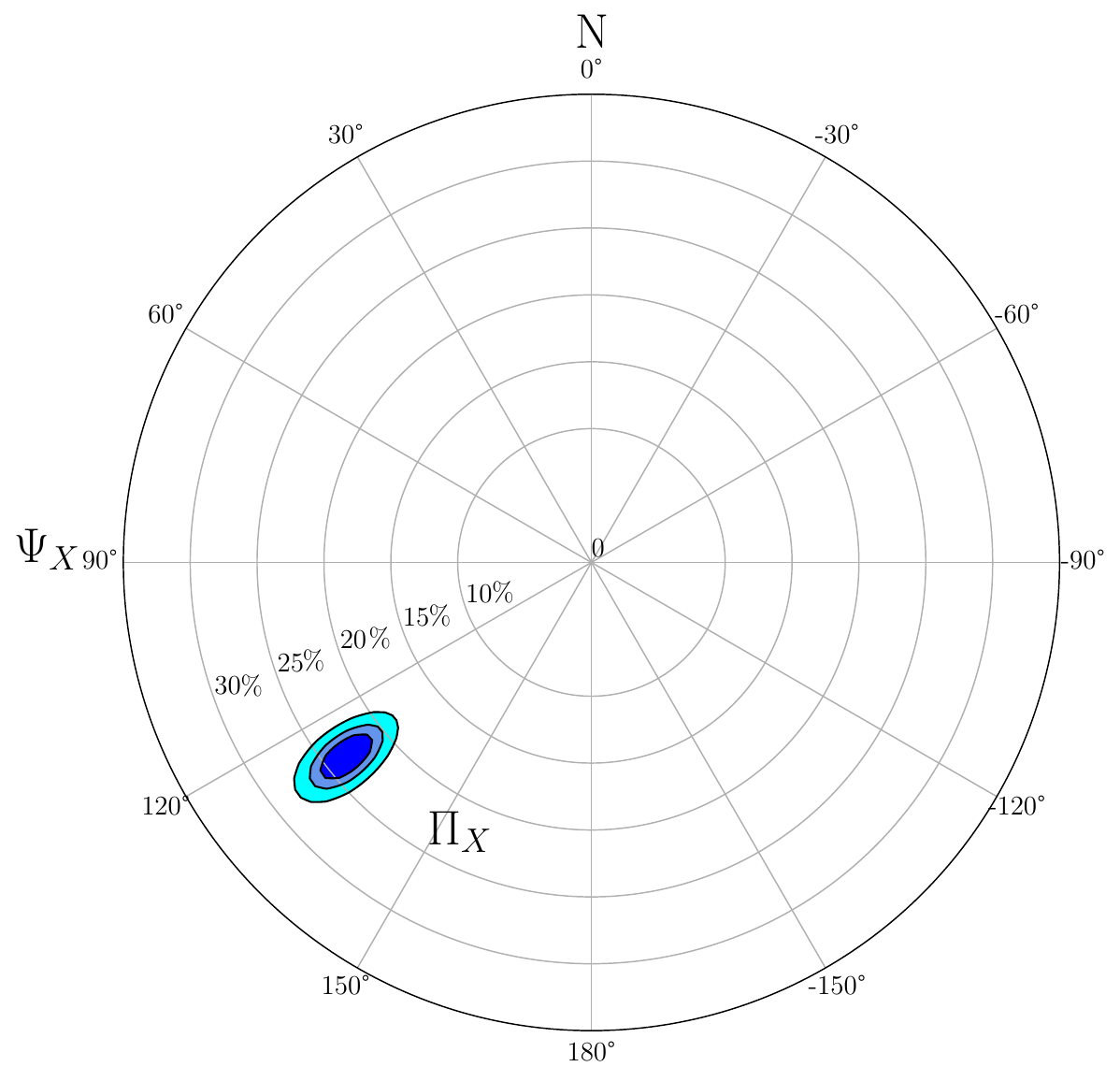}
    \caption{Confidence regions for the time-averaged polarization degree ($\Pi_\mathrm{X}$) and angle ($\Psi_\mathrm{X}$) obtained from the joint IXPE and \textit{XMM-Newton} fit. The contours are shown at 68\%, 90\%, and 99\% confidence levels.}
    \label{poldegangle}
\end{figure}

The results obtained from the spectro-polarimetric modelling are in agreement with a typical HSP spectrum with $\alpha$ in the range 2.55\textendash2.7 and $\beta$=0.078$\pm$0.012. We find that both $\Pi_\mathrm{X}$ and $\Psi_\mathrm{X}$ are significantly constrained, with values of (23.3$\pm$1.5)\% and 128$\degree$$\pm$1.8$\degree$, respectively, in agreement with \cite{Hu2024_pks2155}. In Figure~\ref{poldegangle}, we show the corresponding confidence regions derived for these two parameters. We do not see significant flux variability during the IXPE pointing (Figure~\ref{swift_plot}). Nevertheless, we performed the analysis using the IXPE and \textit{XMM-Newton} datasets separately. We do not find any difference in the derived polarization parameters.

\subsection{Time- and energy-resolved analysis}\label{sec:variability}
\begin{figure}
    \centering
    \includegraphics[width=0.49\textwidth]{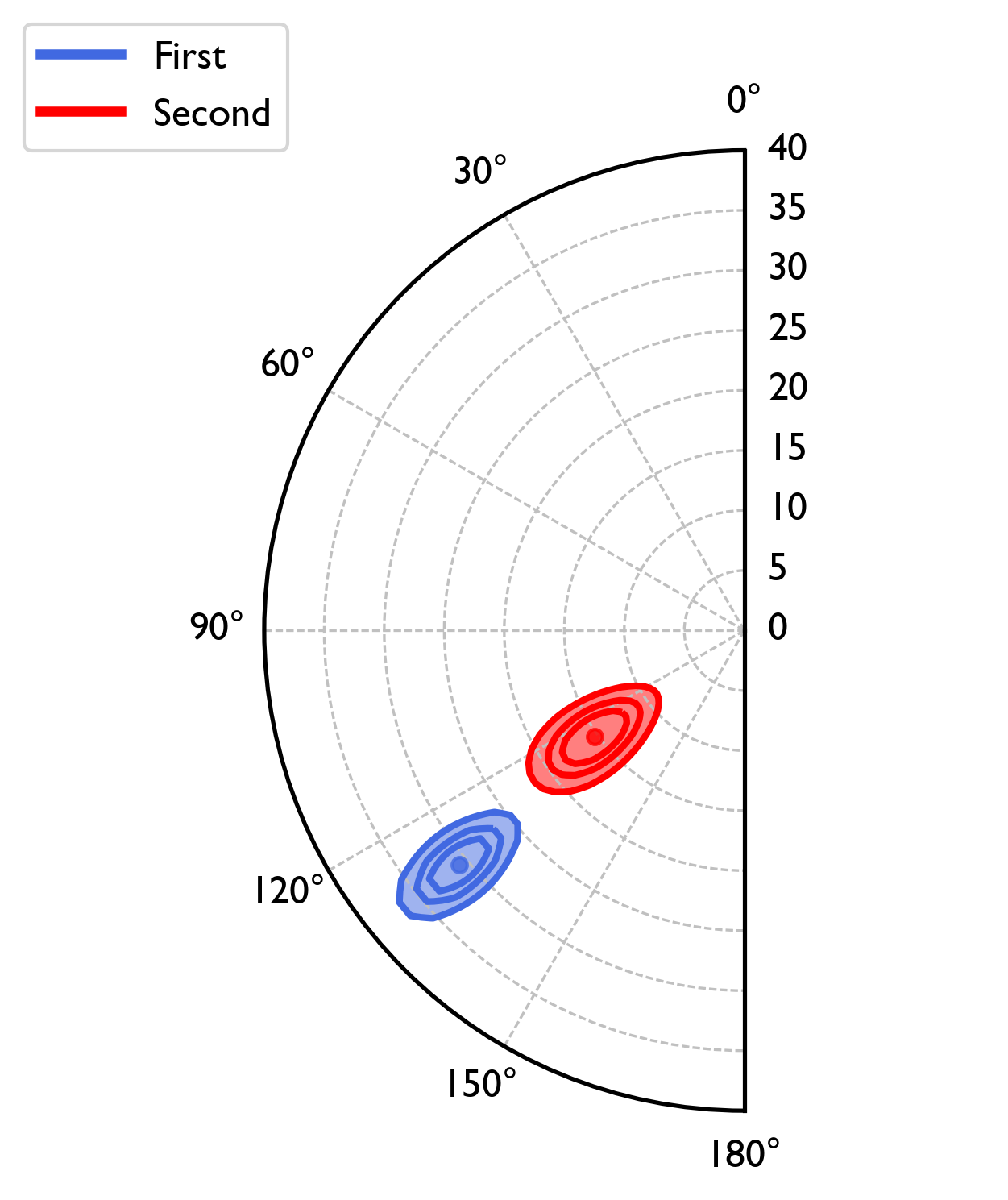}
    \caption{Polarization contours for time periods $T_1$ (blue) and $T_2$ (red), obtained when performing a time-resolved analysis. The contours are shown at confidence levels of 68.27\%, 90.00\%, and 99.00\%.}
    \label{pol_com}
\end{figure}

We additionally investigated the possibility of polarization variability over time and energy using the $\chi^2$ test for a fit of a constant model to the normalized q~=~Q/I and u~=~U/I Stokes parameters, as implemented in \cite{DiGesu2023} and \cite{Kim2024}. We calculated the null-hypothesis probability of the constant model considering the degrees of freedom and $\chi^2$ values from the result of each fit. First, time variability was tested for by dividing the entire observation period into 2\textendash15 sub-periods. For all subdivisions, except 11, 13, and 15, we found that the null-hypothesis probability of the u Stokes parameter being constant throughout the observation was <1\%. The lowest null-hypothesis chance probability found was <0.000127, which occurred when dividing the IXPE pointing into two equal time bins. As shown in Table \ref{time-resolved_Table}, we obtained $\Pi_\mathrm{X}$=(30.7$\pm$2.0)\%, $\Psi_\mathrm{X}$=129.4\degree$\pm$1.8\degree\, for the first period, $T_1$, and $\Pi_\mathrm{X}$=(15.3$\pm$2.1)\%, $\Psi_\mathrm{X}$=125.4\degree$\pm$3.9\degree\, for the second period, $T_2$, as shown in Figure~\ref{pol_com}. The polarization angle was essentially stable, within the 1$\sigma$ uncertainty, from $T_1$ to $T_2$, while the polarization degree varied markedly between the two time bins.

The above time-resolved analysis included data over the entire 2\textendash8~keV IXPE high-sensitivity energy range. We also performed an energy-resolved analysis integrated in time over the entire IXPE pointing. We subdivided the 2\textendash8~keV sensitivity range into multiple energy bins: two 3~keV bins (2\textendash5~keV and 5\textendash8~keV) up to twelve 0.5~keV bins (2.0\textendash2.5~keV, \ldots, 7.5\textendash8.0~keV). A $\chi^2$ test finds no significant difference in polarization properties across the subdivided energy bins. Additionally, we performed a test where the energy and time are divided simultaneously ($T_1$ with 2\textendash5~keV, $T_1$ with 5\textendash8~keV, $T_2$ with 2\textendash5~keV, and $T_2$ with 5\textendash8~keV), we found that there is no significant difference in polarization properties across the subdivided energy bins within the uncertainty of each polarization measurement.

\begin{table}
\setlength{\tabcolsep}{1.9pt}
\centering
\caption{Time-resolved IXPE polarization results of PKS~2155$-$304.\label{time-resolved_Table}}
	\begin{tabular}{c c c}
	\hline\hline
	~Period~ & ~Polarization degree ($\Pi_\mathrm{X}$)~ & ~Polarization angle ($\Psi_\mathrm{X}$)~  \\
	\hline\hline
    $T_1$ & (30.7$\pm$2.0)\% & 129.4\degree$\pm$1.8\degree \\
    $T_2$ & (15.3$\pm$2.1)\% & 125.4\degree$\pm$3.9\degree \\
    \hline
	\end{tabular}
\end{table}

\begin{figure*}
    \centering
    \includegraphics[width=\textwidth]{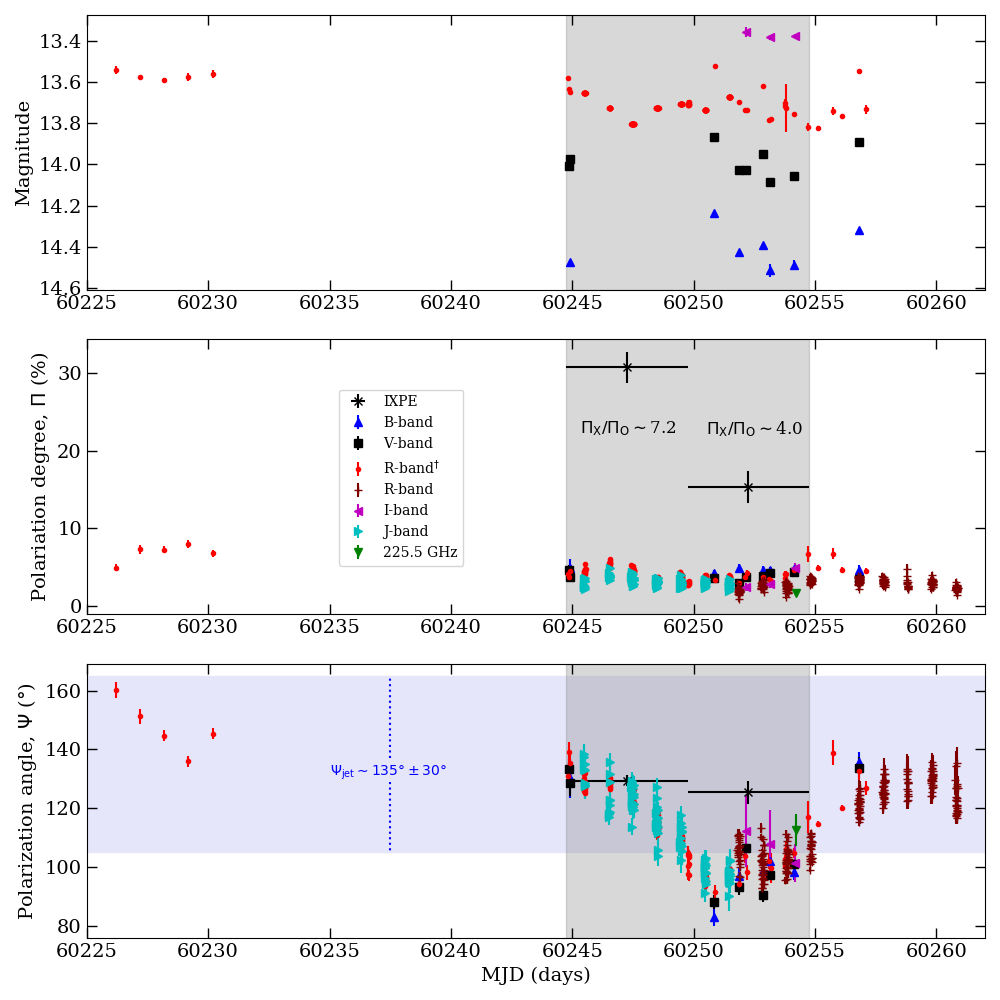}
    \caption{IXPE and contemporaneous multiwavelength polarization of PKS~2155$-$304. Panels from top to bottom are optical brightness, multiwavelength polarization degree, and multiwavelength polarization angle. Vertical gray shaded area demarcates the duration of the IXPE observation. Horizontal lavender shaded area in the bottom panel denotes the approximate position angle of the extended (>~1.5~pc) VLBI jet (see \S\ref{sec:intro}). The symbol $^\dag$ refers to host-galaxy-corrected values. To calculate $\Pi_\mathrm{X}$/$\Pi_\mathrm{O}$, we used the average host-galaxy-corrected $\Pi_\mathrm{O}$ in each of the two IXPE time bins.}
    \label{results_all}
\end{figure*}

\begin{figure}
    \centering
    \includegraphics[width=8cm]{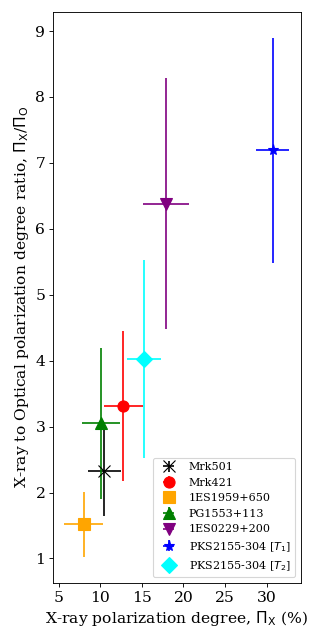}
    \caption{The X-ray to optical polarization degree ratio ($\mathrm{\Pi_X}$/$\mathrm{\Pi_O}$) of the six HSP blazars observed by IXPE plotted against their X-ray polarization degree ($\mathrm{\Pi_X}$). In the case of PKS~2155$-$304, the two time bins ($T_1$ and $T_2$) are presented separately (see \S \ref{sec:variability}), while for the others the detected (>99.7\% confidence) values are averaged over the IXPE pointing.}
    \label{ixpe_all_XvO}
\end{figure}

\begin{figure}
    \centering
    \includegraphics[width=8cm]{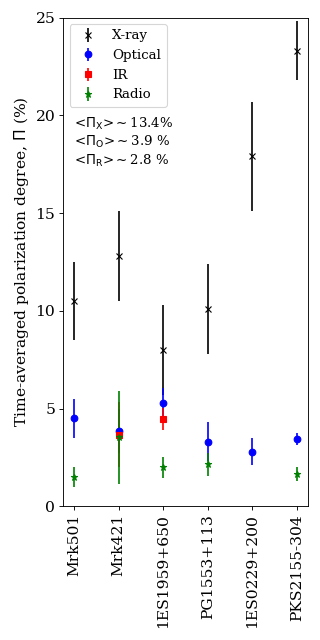}
    \caption{Multiwavelength time-averaged polarization degree of HSP blazars observed by IXPE. Only polarization detections (>99.7\% confidence) were used to calculate the $\Pi$ time-averages.}
    \label{ixpe_all}
\end{figure}

\section{Contemporaneous multiwavelength observations}\label{sec:mwl}
Contemporaneous to the IXPE observation, PKS~2155$-$304 was observed in radio and optical bands. Here we provide a brief description of the observations and data analysis procedures. More details about data reduction can be found in \cite{Liodakis2022nature} and \cite{DiGesu2023}.

At radio frequencies, observations were provided by the Effelsberg 100-m radio telescope and the Korean VLBI Network (KVN). Effelsberg observations were performed within the QUIVER (Monitoring the Stokes Q, U, I and V Emission of AGN jets in Radio) program at 4.85, 10.45, and 13.8~GHz. The KVN observations were performed using the Yonsei and Tamna antennas in single-dish mode \citep{Kang2015} at 25, 43, 86, and 129~GHz. Millimeter-wave radio (mm-radio) observations were provided by the SubMillimeter Array (SMA)  polarimeter (\citealt{Marrone2008}) within the SMAPOL (SMA Monitoring of AGNs with POLarization) program at 225.5~GHz (Myserlis et al. in preparation).

We have found that the source was weakly polarized across all radio bands. At frequencies lower than 129~GHz, we are only able to obtain upper limits. In the 4.85\textendash13.8~GHz regime, the polarization degree is $<3\%$ (99.7\% C.I.). At KVN frequencies, the most constrained upper limit estimate is $<4.4\%$ (99.7\% C.I.), found at 43~GHz. At 225.5~GHz, we observed the source on 2023 October 27, November 4, and November 6. The first two observations yielded 99.7\% C.I. upper limits of $<2.61\%$ and $<0.81\%$. In the third observation, we measured the radio (225.5~GHz) polarization degree to be $\Pi_\mathrm{R}$=(1.7$\pm$0.4)\%, along a polarization angle of $\Psi_\mathrm{R}$=112.5\degree$\pm$5.5\degree. The polarization results in the radio band are summarized in Table \ref{Table_radio_results}.

We conducted contemporaneous optical observations at the Calar Alto Observatory using the Calar Alto Faint Object Spectrograph (CAFOS), the 60~cm telescope at Belogradchik Observatory, the Kanata telescope using the Hiroshima Optical and Near-InfraRed camera (HONIR, \citealp{kawabata_new_1999,akitaya_honir_2014}), the Liverpool Telescope using the Multicolour OPTimised Optical Polarimeter (MOPTOP), the Boston University Perkins Telescope Observatory using the PRISM instrument, and the South African Astronomical Observatory using the HIgh-speed Photo-POlarimeter (HIPPO, \citealp{Potter2010}). HIPPO uses two contra-rotating 1/2 and 1/4 wave-plates and can simultaneously measure linear and circular polarization by fitting the amplitude and phases of the 4th, 8th (linear), and 6th (circular) harmonic \citep{Potter2008,Potter2010}. The 60~cm telescope at Belogradchik observatory uses a set of three polarizing filters, oriented at 0, 60, and 120 degrees, and standard photometric procedures. Details on the analysis and calibration for these observations can be found in \cite{Bachev2023}. MOPTOP features a dual-beam configuration, with a pair of fast-readout, very low-noise CMOS cameras, and a continuously rotating half-wave plate. These allow for high-sensitivity observations while minimizing systematic errors. MOPTOP has a FOV of $7\times7$~arcsec$^2$ (\citealt{Shrestha2020}). Quasi-simultaneous observations were taken in filters B~(380\textendash520~nm), V~(490\textendash570~nm), and R~(580\textendash695~nm). The observations were carried out in slow mode with $16\times4$~s integrations per camera being used to calculate four sets of Stokes IQU parameters. These were averaged before calculating the photopolarimetric data for further minimization of the uncertainties. The photometric data were calculated via standard differential photometry techniques with the \texttt{astropy} and \texttt{photutils} packages in Python, using a calibration star with known BVR magnitudes. The polarimetric data were calibrated using zero-polarized and polarized standard stars to characterize the instrumental polarization, position angle, and depolarization values.

In HSP sources the unpolarized starlight from the host-galaxy has a depolarizing effect on the raw measurements of the optical polarization degree ($\Pi_\mathrm{O}$). To correct for this, the host-galaxy flux density ($I_\mathrm{h}$) contribution within the aperture used in the polarization measurements is needed. We have carefully estimated $I_\mathrm{h}$ for different aperture sizes by convolving the known surface brightness measurement of PKS~2155$-$304 (14.8 R-band magnitudes, equivalent to 3.70~mJy, at an effective radius of 4.5~arcsec; \citealt{Falomo1991_gal_brightness_contour}) to different seeing and subtracting the blazar contribution from it. The modeled host-galaxy magnitude ($m$) against the aperture radius ($r$) can be mathematically represented as $m = \mu_e - 5\log(R_e) - 2.5 \log [2 \pi n \cdot e^{b_n} / (b_n)^{2n} \cdot \Gamma(2n,x) \cdot \gamma(2n)]$, where $x$=$b_n(r/R_e)^{-n}$, $\Gamma$ is the incomplete gamma function, and $\gamma$ is the gamma function (e.g., \citealt{Graham2005_galaxy_brightness_radial_profile}). For the host-galaxy of PKS~2155$-$304, we obtained the fit parameters $n$=1.507, $b_n$=2$n-$0.324, $R_e$=2.87, and $\mu_e$=21.03. For example, when using an aperture of 7.5 arcsec in radius, the host-galaxy flux density contribution is $1.27\pm0.13$~mJy in the R-band. The intrinsic polarization degree ($\Pi_\mathrm{i}$) is then estimated as $\Pi_\mathrm{i}= \Pi_\mathrm{O}\cdot{I_t}/(I_t-I_\mathrm{h})$, where $I_t$ is the total flux density (\citealt{Hovatta2016}). This approach allows for the host-galaxy-correction of all R-band polarization measurements that are accompanied by a photometric image. This was performed for several R-band data sets, whose host-galaxy-corrected $\Pi_\mathrm{O}$\footnote{Throughout this paper the subscripts O and R refer to the generic optical (including near-IR) and radio bands, respectively. Any filter specific information then follows the generic subscript in parentheses.} values are shown in Figure~\ref{results_all}, labeled as R-band$^\dag$. We note that the host-galaxy contribution is dependent on the wavelength. For example, the host-galaxy is brighter in IR bands, while dimmer in higher energy optical bands. Unfortunately, the aforementioned modeling of the host-galaxy was only possible in the R-band. Therefore, we can only correct for the host-galaxy contribution in the R-band.

\begin{table}
\setlength{\tabcolsep}{1.9pt}
\centering
\caption{Linear polarization results of PKS~2155$-$304 in the radio bands.\label{Table_radio_results}}
	\begin{tabular}{r l c l l}
	\hline\hline
	Program & Frequency (GHz) & MJD & $\Pi_\mathrm{R}$ & $\Psi_\mathrm{R}$  \\
	\hline\hline
    QUIVER & 4.85 & 60254 & $<2.13\%$ & --- \\
    QUIVER & 10.45 & 60254 & $<3.33\%$ & --- \\
    QUIVER & 13.85 & 60254 & $<2.76\%$ & --- \\
    KVN & 23 & 60261 & $<8.8\%$ & --- \\
    KVN & 43 & 60261 & $<4.4\%$ & --- \\
    KVN & 86 & 60261 & $<16.6\%$ & --- \\
    KVN & 129 & 60261 & $<13.0\%$ & --- \\
    SMAPOL & 225.5 & 60244 & $<2.6\%$ & --- \\
    SMAPOL & 225.5 & 60252 & $<0.8\%$ & --- \\
    SMAPOL & 225.5 & 60254 & (1.7$\pm$0.4)\% & 112.5\degree$\pm$5.5\degree \\
    \hline
	\end{tabular}
    \tablefoot{The upper limit values of $\Pi_\mathrm{R}$, shown using "$<$", correspond to 99.7\% C.I.}
\end{table}

Table \ref{Table_optical_obs} summarizes the optical observations of PKS~2155$-$304. During the IXPE pointing (between MJD 60244 and 60255), we find that $\Pi_\mathrm{O}$ was rather stable, with average host-galaxy-corrected R-band values of (4.3$\pm$0.7)\% and (3.8$\pm$0.9)\% for the first and second halves of the pointing, respectively. We measure $\Psi_\mathrm{O}$ to be in the 90\degree\textendash140\degree range. We note that while we were only able to perform host-galaxy-correction in the R-band (which is rather narrow), the host-galaxy contribution in the B and V bands is usually negligible. In the I and J bands the contribution is expected to be greater, which would shift their corrected polarization degrees closer to the corrected R-band values. Due to the high density of data points in the optical and near-IR bands (see Figure~\ref{results_all}), in Appendix \ref{appendix:detailed_opt_lc} (Figure~\ref{detailed_opt_plot}) we plotted zoomed-in light curves to allow for a better analysis of the color-dependent properties of these bands. We find that while $\Pi_\mathrm{O}$ underwent rather achromatic variations, the simultaneous polarization degrees at B-band were consistently higher than those of the host-galaxy-corrected R-band, with an average ratio around 1.2 (see Figure~\ref{detailed_opt_plot}). In Appendix \ref{appendix:detailed_opt_lc}, we additionally visualize the chromatic intra-night variations observed during the IXPE pointing. The polarization angle $\Psi_\mathrm{O}$ exhibited a continuous and achromatic rotation from $\sim$140\degree\, to $\sim$90\degree\, between MJD 60244 and 60251, after which the value increased to $\sim$130\degree\, (at MJD 60257 and staying roughly the same during the following three nights). While the average source brightness differed strongly with color in the optical bands, with the source being brighter in the redder bands due to stronger host-galaxy contamination, the time-resolved brightness of all the colors showed little variability during the IXPE pointing. Overall, these values were comparable to the typical ones seen in the decade-long monitoring results of PKS~2155$-$304 at Steward observatory (see Appendix \ref{appendix:longterm}), indicating that the source was in an average optical state when IXPE observed it. We note that $\Pi_\mathrm{O}$ was a factor of $\sim$2 greater than the polarization degree in the radio band ($\Pi_\mathrm{R}$) and $\Psi_\mathrm{O}$ was consistent with the simultaneous radio band polarization angle ($\Psi_\mathrm{R}$); see Figure~\ref{results_all}.

Optical (R-band) circular polarization observations of the source using HIPPO yielded an upper limit of <0.87\% at 99.7\% C.I. during the IXPE observation.

\begin{table}
\setlength{\tabcolsep}{1.9pt}
\centering
\caption{Linear polarization results of PKS~2155$-$304 in the optical and near-IR bands.\label{Table_optical_obs}}
	\begin{tabular}{r l c c r c c c c}
    \hline\hline 
    O & Instr. & First & Last & \# & $\overline{\mathrm{m}}$ & $\overline{\Pi}$ & $\widetilde{\sigma_\Pi}$ & $\overline{\Psi}$ \\ 
    \hline\hline 
    B & PRISM & 60253 & 60254 & 2 & 14.5 & 4.6\% & $\pm$0.4\% & 100.2\degree \\ 
    B & MOPTOP & 60244 & 60256 & 5 & 14.4 & 4.7\% & $\pm$0.5\% & 108.9\degree \\ 
    \hline 
    V & PRISM & 60252 & 60254 & 3 & 14.1 & 4.1\% & $\pm$0.2\% & 101.5\degree \\ 
    V & MOPTOP & 60244 & 60256 & 6 & 14.0 & 3.7\% & $\pm$0.4\% & 111.2\degree \\ 
    \hline 
    R & PRISM & 60226 & 60257 & 13 & 13.7 & 5.2\%$^\dag$ & $\pm$0.4\% & 123.7\degree \\ 
    R & MOPTOP & 60244 & 60256 & 7 & 13.6 & 3.5\%$^\dag$ & $\pm$0.4\% & 117.1\degree \\ 
    R & HONIR & 60245 & 60251 & 109 & 13.7 & 4.1\%$^\dag$ & $\pm$0.2\% & 113.7\degree \\ 
    R & CAFOS & 60249 & 60253 & 11 & 13.7 & 3.2\%$^\dag$ & $\pm$0.2\% & 101.9\degree \\ 
    R & Belogr. & 60254 & 60255 & 2 & 13.8 & 6.7\%$^\dag$ & $\pm$0.9\% & 127.9\degree \\ 
    R & HIPPO & 60251 & 60260 & 254 & --- & 2.8\% & $\pm$0.3\% & 114.7\degree \\ 
    \hline 
    I & PRISM & 60252 & 60254 & 3 & 13.4 & 3.4\% & $\pm$0.6\% & 107.2\degree \\ 
    \hline 
    J & HONIR & 60245 & 60251 & 99 & --- & 3.1\% & $\pm$0.4\% & 113.4\degree \\ 
    \hline 
    \end{tabular}
    \tablefoot{"O" represents the optical band of the measurement. "Instr." represents the instrument used to make the measurement. Note that "Belogr." refers to Belogradchik Observatory. "First" and "Last" represent the first and last MJDs on which data from the given instrument was recorded, respectively. "\#" represents the number of data points taken via each instrument within the given the time window. "$\overline{\mathrm{m}}$", "$\overline{\Pi}$", "$\widetilde{\sigma_\Pi}$", and "$\overline{\Psi}$" give the average magnitude, average polarization degree, median standard deviation of the polarization degree, and average polarization angle within the time window, respectively. Note that some $\overline{\Pi}$ values are host-galaxy corrected (represented by the symbol $^\dag$).}
\end{table}

\section{Discussion}\label{sec:disc}
Several particle acceleration mechanisms have been suggested to explain how electron populations in HSP blazar jets can reach the energies needed for their synchrotron luminosity to peak in the X-ray regime. Diffusive shock acceleration in weakly magnetized jets is one such mechanism (e.g., \citealt{Blandford1987_shock_acc}). For example, \cite{Marscher2014_multi_zone_shock} considered a scenario where a conical standing shock energizes turbulent plasma as it crosses the shock front, with some regions being accelerated more effectively due to favorable orientation of their magnetic field lines relative to the front. The particles lose energy by radiating as they advect farther from the shock, hence there is a gradient in maximum particle energy in the emitting region. This multi-zone emission model can explain the general nature of temporal variations observed in the flux and polarization. In the case of an HSP blazar, this model predicts a higher mean value of $\Pi_\mathrm{X}$ with higher-amplitude variations than at longer wavelengths (\citealt{marscher2022_theoretical_estimate, Peirson2018_mw_pol_behavior_synch_jet}). It additionally predicts random rotations of $\Psi$ with varying rate, magnitude $\Delta\Psi$, and direction (clockwise or counterclockwise). 

Particle acceleration in a shock or compressed plasmoid moving down a jet with a partially ordered or helical magnetic field has also been suggested (e.g., \citealt{Blandford1979,Marscher1985_shock_wave,Sikora1994,Marscher2008_nature_MW_EVPA_rot}). If the plasmoid has uniform physical conditions, the multi-frequency emission is co-spatial, hence the polarization patterns across different frequencies are expected to be similar (\citealt{DiGesu2022_time_var}). In contrast, a moving shock should have similar multiwavelength properties as described above for a standing shock, with stratification of the maximum energy, and therefore frequency, profile. The volume occupied by the lower-energy particles is greater, hence vector averaging of the disordered, or otherwise multi-directional, magnetic field lowers the polarization at longer wavelengths. This leads to wavelength dependence of $\Pi$, whose mean value is expected to decrease toward longer wavelengths (e.g., \citealt{Angelakis2016,Tavecchio2018_shock_enegy_stratified_prediction}). Furthermore, as the flow crosses the shock front, the component of the internal magnetic field that is parallel to the shock front becomes more ordered, which results in $\Psi$ at the synchrotron peak frequency ($\Psi_\mathrm{X}$ for HSP blazars) to align with the jet direction. In the case of a helically twisted magnetic field structure in the jet (evidence for which is found in parsec-scale VLBI observations; e.g., \citealt{Hovatta2012_mojave_helical}), $\Psi$ at the synchrotron SED peak frequency is expected to exhibit non-stochastic rotations resulting from the passage of the shock front through the jet. In blazars whose SED peaks in the optical regime, systematic $\Psi_\mathrm{O}$ rotations have been observed, with some being temporally correlated with GeV $\gamma$-ray flares (\citealt{Blinov2015_robopol_evpa_rot_p1,Blinov2018_robopol_evpa_rot_p2}). Additionally, a harder-when-brighter spectral behavior can be expected in the case of shock front acceleration (\citealt{Kirk1998_shock_harder_when_brighter}).

If the jets are highly magnetized, shock acceleration is not effective. Instead, magnetic reconnection could efficiently convert magnetic energy into particle energy (\citealt{Sironi2014_reconn_in_mag_jets, Sironi2015_reconn_in_mag_jets}). For example, this can involve current sheets generated by kink instabilities  (\citealt{Bodo2021_kink_instability}). Further stochastic acceleration can follow injection of particles into turbulent regions (\citealt{Comisso2018_turbulent_acc}). The expected observed polarization resulting from magnetic reconnection scenarios, averaged over the IXPE $\sim$10\textendash100~ks integration times, would correspond to $\Pi_\mathrm{X}$$\lesssim$$\Pi_\mathrm{O}$ and a different temporal evolution for $\Psi_\mathrm{X}$ than $\Psi_\mathrm{O}$ (\citealt{DiGesu2022_time_var}).

Alternatively, \cite{Zhang2020_striped_jet_mag_reconn} considered a striped-jet scenario where more antiparallel magnetic field lines are expected to form than in the kink instability/turbulent scenarios, leading to the production, merger, and acceleration of plasmoids. Their observational predictions include a harder-when-brighter behavior and more temporal variability of $\Pi_\mathrm{X}$ than $\Pi_\mathrm{O}$ as $\Psi_\mathrm{X}$ undergoes significant 90\degree\textendash180\degree\, rotations along the anti-parallel field lines. Integrated over an IXPE pointing, this is expected to yield $\Pi_\mathrm{X}$<$\Pi_\mathrm{O}$.

In the case of PKS~2155$-$304, we find that $\Psi$ at all wavelengths is in the general direction of the parsec-scale jet seen in the VLBI images (which is generally southeast at $\sim$135\degree; see \S\ref{sec:intro}). Our single significant radio detection of polarization measured $\Psi_\mathrm{R}\sim$110\degree, which is roughly aligned with the jet direction. In the X-ray band, we find no evidence that $\Psi_\mathrm{X}$ changed during the IXPE pointing, while in the optical band it rotated from $\Psi_\mathrm{O}\sim$140\degree\, at the beginning to $\sim$90\degree\ at around MJD 60251, then up to $\sim$100\degree\ at the end of the IXPE pointing, and further up to $\sim$130\degree at MJD 60257. We note that such $\sim\pm25$\degree\, fluctuations about a mean value of $\Psi_\mathrm{O}$ is typical of PKS~2155$-$304 as tracked by Steward observatory (see Appendix \ref{appendix:longterm}). However, the mean value gradually drifted from $\sim80$\degree\, to $\sim40$\degree\, in around 10 years ending in 2020. This is quite different in both rate and range of values from the rotation on a timescale of days seen in our data set, throughout which $\Psi$ at all wavelengths is roughly aligned with the VLBI jet direction. This is a strong indication of particle acceleration occurring in regions with a magnetic field predominantly oriented orthogonally to the jet axis, as expected in the shock acceleration scenario or for a jet with a dominant toroidal field. Additionally, the discrepant behavior between $\Psi_\mathrm{X}$ and $\Psi_\mathrm{O}$ indicates that at least part of their emission arises in separate regions in the jet. Furthermore, the lack of evidence for any fast changes in $\Psi$ strongly disfavors particle acceleration mechanisms resulting from magnetic reconnection or large-scale turbulence.

Regarding the polarization degree, we find that $\Pi_\mathrm{X}$>$\Pi_\mathrm{O}$>$\Pi_\mathrm{R}$ held true throughout the IXPE pointing, with $\Pi_\mathrm{X}$/$\Pi_\mathrm{O}$ being $\sim$7.2 and $\sim$4.0 in the first and second halves of the pointing, respectively (see Figure~\ref{results_all}). The former (during $T_1$) is the most extreme X-ray-to-optical polarization degree ratio found in a blazar to date, as seen in Figure~\ref{ixpe_all_XvO} (also see Figure~\ref{ixpe_all}): more than twice that measured for Mrk~421, Mrk~501, and 1ES~1959+650. Moreover, we note that PKS~2155$-$304 has historically had the softest X-ray spectrum as compared with the others (according to the ROSAT all-sky survey; \citealt{Boller2016_ROSAT_survey}), which held true during the IXPE pointing. A softer X-ray spectrum is indicative of the X-ray emission originating farther above the synchrotron peak frequency. The particles radiating at X-ray energies in PKS~2155$-$304 have shorter cooling times and therefore occupy smaller volumes than in the other HSP blazars, according to our interpretation. Emission from such particles would result in a higher mean value --- as well as higher-amplitude variability --- of $\Pi_\mathrm{X}$, as the averaging effects due to randomly-oriented magnetic fields would be reduced (\citealt{Peirson2023}). On the other hand, being farther above the synchrotron peak frequency may result in contribution by the high-energy component of the SED in the 2\textendash8~keV band (e.g., \citealt{Madejski2016_pks2155_IC_contribution}). If that contribution is significant, it may result in somewhat underestimating the measured $\Pi_\mathrm{X}$.

Interestingly, the more-polarized-at-higher-frequencies behavior, observed when comparing the X-ray to the optical band results, also occurred within the optical band: the average ratio of the simultaneous B-band to host-galaxy-corrected R-band polarization degrees was 1.2 (see Appendix \ref{appendix:detailed_opt_lc}). As discussed above and more extensively in \cite{Liodakis2022nature} and \cite{DiGesu2022_time_var}, for example, this behavior is expected in the scenario where a compact shock front accelerates particles and partially orders the magnetic field. However, the ratio is substantially larger than predicted by the basic model (see Appendix \ref{appendix:detailed_opt_lc}).

Notably, we find that, while there were no temporal variations in the optical polarization degree $\Pi_\mathrm{O}$, there was a significant drop in the X-ray polarization degree $\Pi_\mathrm{X}$ during the IXPE pointing. This reaffirms that the X-ray and longer-wavelength emissions at least partly arose from different regions of the jet. It also indicates that the magnetic field in the X-ray emission region became significantly more disordered over a few days, despite its average orientation remaining constant, as indicated by the constant polarization angle. The X-ray flux also remained steady within the uncertainties; see Appendix \ref{appendix:longterm}. Such a decrease in order of the magnetic field lines could be caused by a significant strengthening of turbulence in the close vicinity of the shock front (where the high-energy electrons emitting in the X-ray band lose their energy; \citealt{Angelakis2016}). As discussed in \cite{Liodakis2022nature} and \cite{marscher2022_theoretical_estimate}, the polarization of the turbulent component of the field is expected to vary with an amplitude equal to the square-root of the mean polarization degree. While this could explain the change in $\Pi_\mathrm{X}$ from $T_1$ to $T_2$, a strong variation of $\Psi_\mathrm{X}$ would also be expected, but was not observed. Instead, as mentioned above, the emission region during $T_1$ may have been relatively small, i.e., a few acceleration zones were dominating, resulting in a high value of $\Pi_\mathrm{X}$. If more acceleration zones were boosted into the emission range detectable by IXPE, then a drop in the mean $\Pi_\mathrm{X}$ during $T_2$ would be expected.

Alternatively, in the framework of the stratified shock model of \cite{Tavecchio2018_shock_enegy_stratified_prediction}, a lower polarization degree can be related either to a more rapid decay of the partially ordered magnetic field with distance from the shock (related to the microphysics of the shock) or by a strengthening of the poloidal field carried by the pre-shocked plasma.  Detailed modeling of the multiwavelength data is required to extract more precise information and to effectively test the different theoretical scenarios.

To demonstrate the similarity of the radio through X-ray polarization behavior of the six HSP targets observed by IXPE thus far, we plot their average polarization degree ($\Pi$) at each general wave band from multiwavelength observations in Figure~\ref{ixpe_all}. It is apparent that, on average, $\Pi_\mathrm{X}$ is several times greater than $\Pi_\mathrm{O}$, which is a natural prediction of the energy-stratified shock-acceleration scenario. If this scenario is at play, then the fairly good agreement between the lower energy multiwavelength polarization degrees (<$\Pi_\mathrm{O}$> and <$\Pi_\mathrm{R}$>) over the different HSP blazars implies that, at a given frequency, their emission regions have similar sizes as well as structure/ordering of the magnetic field. However, the significant scatter in the observed X-ray polarization degree (<$\Pi_\mathrm{X}$>) of the HSP blazars suggests that shock properties may be prone to greater variability at higher energies.

\section{Conclusions}\label{sec:conc}
IXPE observations of PKS~2155$-$304 have revealed the highest X-ray polarization yet detected among the six similarly observed HSP blazars. During the $\sim$10-day IXPE pointing, the X-ray polarization degree ($\Pi_\mathrm{X}$) decreased from $\sim$30\% during the first half to $\sim$15\% during the second half. Meanwhile, the optical polarization degree ($\Pi_\mathrm{O}$) remained stable at $\sim$4\%. The X-ray to optical ratios of $\sim$7 and $\sim$4 in the first and second halves, respectively, are similar to the previous results obtained for HSP blazars (summarized in Figure~\ref{ixpe_all}). This consistency disfavors reconnection-based and stochastic turbulent (second-order Fermi) mechanisms for explaining the high energies of particles needed to produce X-ray synchrotron emission in HSP blazars. 

During the IXPE pointing, the polarization angle ($\Psi$) of PKS~2155$-$304 at all monitored wavelengths was roughly aligned with the direction of the parsec-scale jet. This general alignment of the multiwavelength $\Psi$ with the jet direction indicates that the magnetic field in the particle acceleration region was partially ordered along a direction transverse to the jet. While the X-ray polarization angle remained stable during the IXPE exposure, a clear change in the optical polarization angle was observed. This, combined with the lack of temporal correlation between $\Pi_\mathrm{X}$ and $\Pi_\mathrm{O}$, suggests that the X-ray and optical emission regions are not completely co-spatial. These findings, well in line with those of previous studies of HSP polarization involving IXPE, can be qualitatively explained by the energy-stratified shock-acceleration scenario. 
However, the observations executed thus far have sampled only a small range of time intervals, leaving the temporal behavior of the X-ray polarization rather poorly studied.
Future X-ray and multiwavelength polarization measurements of HSP blazars are needed in order to determine whether the particle acceleration and other physical characteristics change with time. 
%

\begin{acknowledgements}
The Imaging X-ray Polarimetry Explorer (IXPE) is a joint US and Italian mission.  The US contribution is supported by the National Aeronautics and Space Administration (NASA) and led and managed by its Marshall Space Flight Center (MSFC), with industry partner Ball Aerospace (contract NNM15AA18C).  The Italian contribution is supported by the Italian Space Agency (Agenzia Spaziale Italiana, ASI) through contract ASI-OHBI-2017-12-I.0, agreements ASI-INAF-2017-12-H0 and ASI-INFN-2017.13-H0, and its Space Science Data Center (SSDC), and by the Istituto Nazionale di Astrofisica (INAF) and the Istituto Nazionale di Fisica Nucleare (INFN) in Italy. This research used data products provided by the IXPE Team (MSFC, SSDC, INAF, and INFN) and distributed with additional software tools by the High-Energy Astrophysics Science Archive Research Center (HEASARC), at NASA Goddard Space Flight Center (GSFC). This work has been partially supported by the ASI-INAF program I/004/11/4. The IAA-CSIC co-authors acknowledge financial support from the Spanish "Ministerio de Ciencia e Innovaci\'{o}n" (MCIN/AEI/ 10.13039/501100011033) through the Center of Excellence Severo Ochoa award for the Instituto de Astrof\'{i}isica de Andaluc\'{i}a-CSIC (CEX2021-001131-S), and through grants PID2019-107847RB-C44 and PID2022-139117NB-C44. The Submillimetre Array is a joint project between the Smithsonian Astrophysical Observatory and the Academia Sinica Institute of Astronomy and Astrophysics and is funded by the Smithsonian Institution and the Academia Sinica. Mauna Kea, the location of the SMA, is a culturally important site for the indigenous Hawaiian people; we are privileged to study the cosmos from its summit. E.L. was supported by Academy of Finland projects 317636 and 320045. The research at Boston University was supported in part by National Science Foundation grant AST-2108622, NASA Fermi Guest Investigator grant 80NSSC23K1507, and NASA \textit{Swift} Guest Investigator grant 80NSSC23K1145. The Perkins Telescope Observatory, located in Flagstaff, AZ, USA, is owned and operated by Boston University.  I.L was funded by the European Union ERC-2022-STG - BOOTES - 101076343. Views and opinions expressed are however those of the author(s) only and do not necessarily reflect those of the European Union or the European Research Council Executive Agency. Neither the European Union nor the granting authority can be held responsible for them. Some of the data are based on observations collected at the Centro Astron\'{o}mico Hispano en Andalucía (CAHA), operated jointly by Junta de Andaluc\'{i}a and Consejo Superior de Investigaciones Cient\'{i}ficas (IAA-CSIC). This work was supported by NSF grant AST-2109127. We acknowledge the use of public data from the \textit{Swift} data archive. Based on observations obtained with \textit{XMM-Newton}, an ESA science mission with instruments and contributions directly funded by ESA Member States and NASA. Partly based on observations with the 100-m telescope of the MPIfR (Max-Planck-Institut f\"ur Radioastronomie) at Effelsberg. Observations with the 100-m radio telescope at Effelsberg have received funding from the European Union’s Horizon 2020 research and innovation programme under grant agreement No 101004719 (ORP). I.L was supported by the NASA Postdoctoral Program at the Marshall Space Flight Center, administered by Oak Ridge Associated Universities under contract with NASA.  S. Kang, S.-S. Lee, W. Y. Cheong, S.-H. Kim, and H.-W. Jeong were supported by the National Research Foundation of Korea (NRF) grant funded by the Korea government (MIST) (2020R1A2C2009003). The KVN is a facility operated by the Korea Astronomy and Space Science Institute. The KVN operations are supported by KREONET (Korea Research Environment Open NETwork) which is managed and operated by KISTI (Korea Institute of Science and Technology Information). This work was supported by JST, the establishment of university fellowships towards the creation of science technology innovation, Grant Number JPMJFS2129. This work was supported by Japan Society for the Promotion of Science (JSPS) KAKENHI Grant Numbers JP21H01137. This work was also partially supported by Optical and Near-Infrared Astronomy Inter-University Cooperation Program from the Ministry of Education, Culture, Sports, Science and Technology (MEXT) of Japan. We are grateful to the observation and operating members of Kanata Telescope. Data from the Steward Observatory spectropolarimetric monitoring project were used. This program is supported by Fermi Guest Investigator grants NNX08AW56G, NNX09AU10G, NNX12AO93G, and NNX15AU81G. This research was partially supported by the Bulgarian National Science Fund of the Ministry of Education and Science under grants KP-06-H38/4 (2019) and KP-06-PN-68/1 (2022). The Liverpool Telescope is operated on the island of La Palma by Liverpool John Moores University in the Spanish Observatorio del Roque de los Muchachos of the Instituto de Astrofisica de Canarias with financial support from the UKRI Science and Technology Facilities Council (STFC) (ST/T00147X/1). We thank Talvikki Hovatta for fruitful discussion about the VLBI images.
\end{acknowledgements}

\bibliographystyle{aa} 
\bibliography{ref.bib} 

\begin{appendix}
\section{Long-term variability of PKS~2155$-$304} \label{appendix:longterm}
\begin{figure*}
    \centering
    \includegraphics[width=16cm]{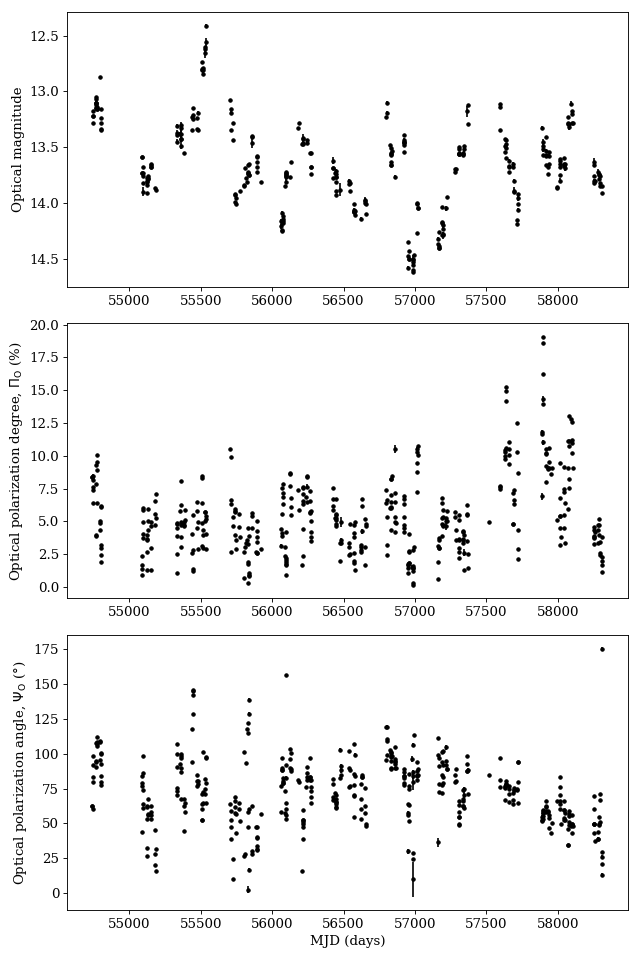}
    \caption{Decade-long flux and polarization monitoring of PKS~2155$-$304 in the optical band at Steward observatory (\citealt{Smith2009_steward_opt_pol_long_term}). \textit{Top:} brightness in magnitudes; \textit{middle:} polarization degree; \textit{bottom:} polarization angle.}
    \label{steward_plot}
\end{figure*}

\begin{figure*}
    \centering
    \includegraphics[width=\textwidth]{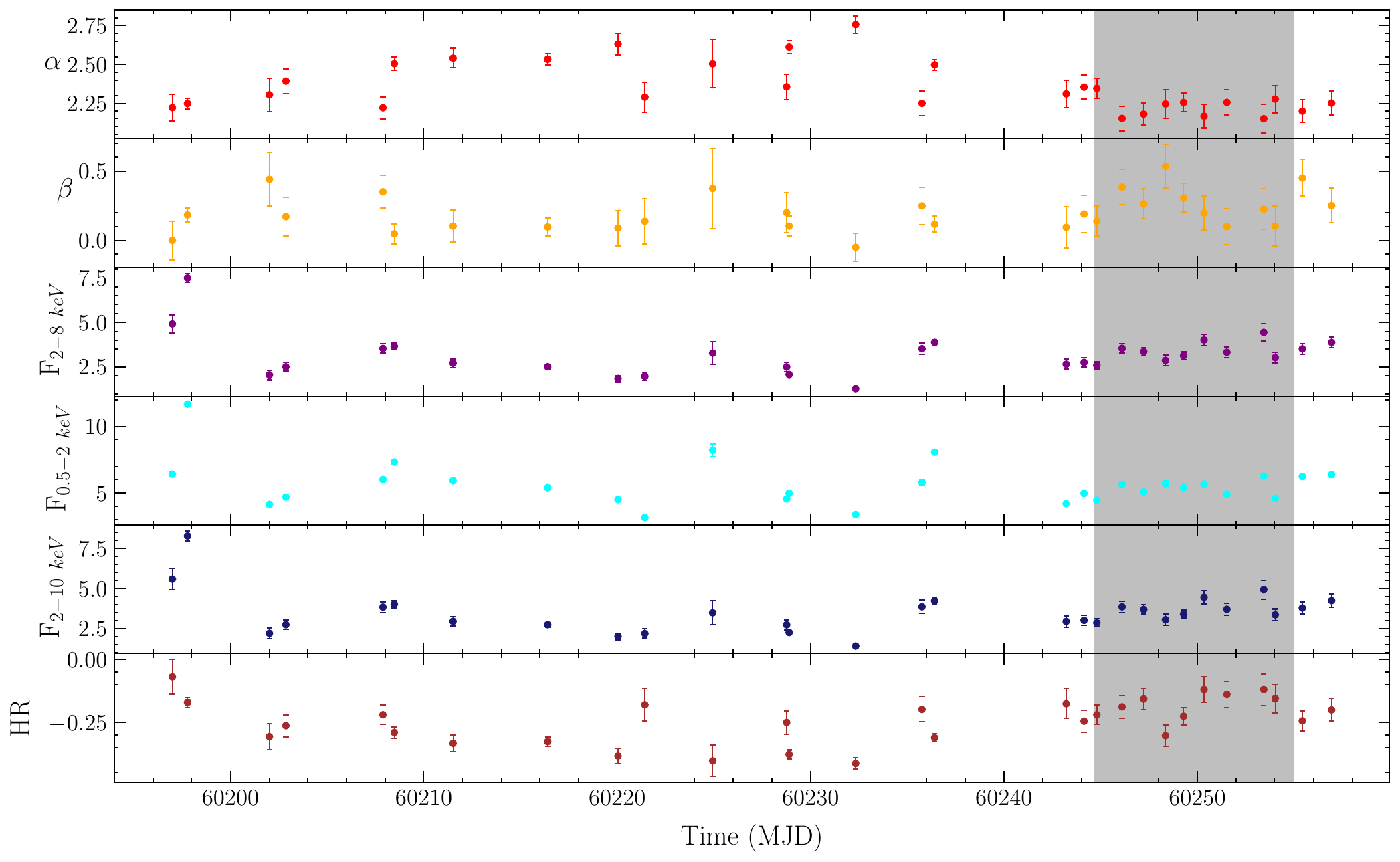}
    \caption{Two-month long \textit{Swift}-XRT X-ray light curve of PKS~2155$-$304. The best-fit parameters of the logarithmic parabola spectral model, as well as the X-ray fluxes (in the units: $10^{-11}$~erg~cm$^{-2}$~s$^{-1}$) in different bands and their ratio (0.5-2/2-10), are shown as a function of time. Gray shaded area identifies epochs of the IXPE observation. PKS~2155$-$304 was in an average X-ray flux state during the IXPE pointing.}
    \label{swift_plot}
\end{figure*}

It is interesting to place the behavior of PKS~2155$-$304 during the IXPE pointing in the context of the longer-term variations. In Figure~\ref{steward_plot} we present the light curve from 10 years of optical polarization monitoring at Steward observatory (\citealt{Smith2009_steward_opt_pol_long_term}). Additionally, in Figure~\ref{swift_plot}, we present the two-month-long X-ray flux versus time as measured by \textit{Swift}-XRT (see \S\ref{sec:xray_pol_swift}).

The Steward optical data clearly show that the average flux state of the source in the V-band is between 13\textendash14 magnitudes. It additionally shows that the optical polarization degree (measured in the wavelength range of 500\textendash700~nm) typically varies between 2\%\textendash10\%, while its polarization angle seems to generally vary between 60\degree\textendash120\degree. The observed optical brightness and polarization properties of PKS~2155$-$304 during the IXPE pointing fall well in line with these typical values (see \S\ref{sec:mwl}). This implies that the source was in an average optical state when IXPE observed it. Likewise, the two-month long \textit{Swift}-XRT data, although much shorter than 10 years, suggests that PKS~2155$-$304 was also in an average flux state in the X-ray band.

\section{Detailed optical and near-IR light curves of PKS~2155$-$304} \label{appendix:detailed_opt_lc}
\begin{figure*}
    \centering
    \includegraphics[width=15cm]{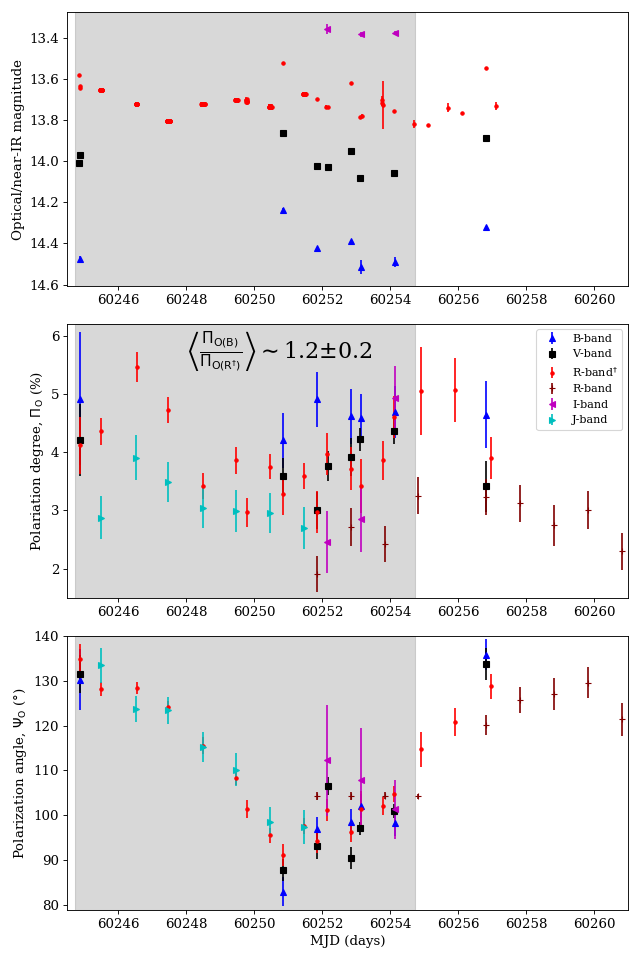}
    \caption{Optical and near-IR light curves of PKS~2155$-$304 during and after the IXPE pointing. Using the largest daily gaps, we binned the data points of each filter. We note that only some of the optical R-band data points are host-galaxy-corrected, which are labeled as R$^\dag$. The average, simultaneous ratio of the polarization degree of the B-band (380\textendash520~nm, i.e., 2.4\textendash3.3~eV), $\Pi_\mathrm{O(B)}$, to the polarization degree of the host-galaxy-corrected R-band (580\textendash695~nm, i.e., 1.8\textendash2.1~eV), $\Pi_\mathrm{O(R^\dag)}$, is around $1.2\pm0.2$. The gray shaded area shows the IXPE pointing time window. \textit{Top:} brightness in magnitudes; \textit{middle:} polarization degree; \textit{bottom:} polarization angle.}
    \label{detailed_opt_plot}
\end{figure*}

\begin{figure*}
    \centering
    \includegraphics[width=17.5cm]{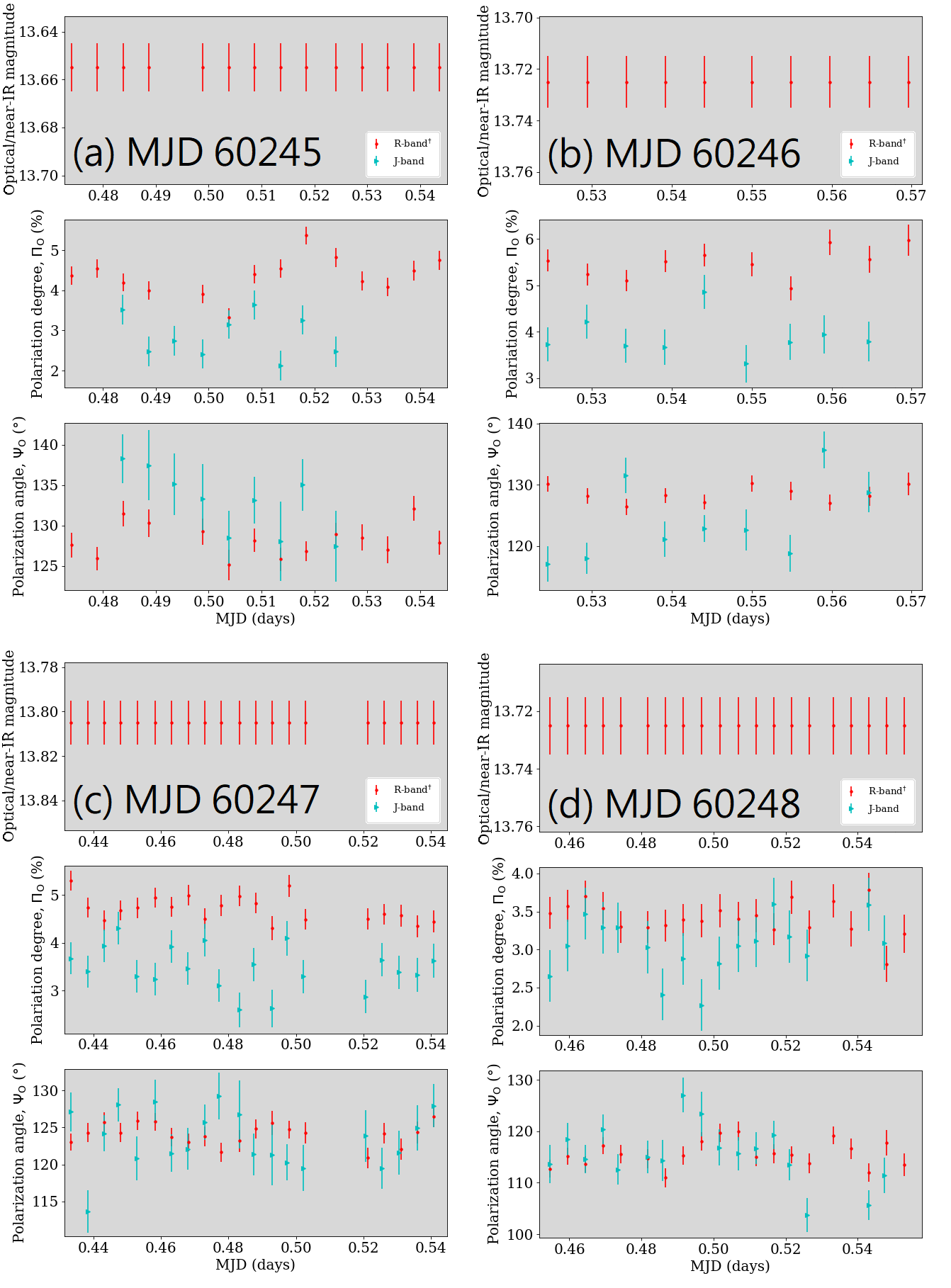}
    \caption{Zoomed-in optical and near-IR light curves of PKS~2155$-$304 on the four indicated example nights. The red data points (R$^\dag$) are those of the host-galaxy-corrected optical R-band, and the cyan ones are those of the near-IR J-band (not host-galaxy-corrected).}
    \label{detailed_opt_plot_nightly}
\end{figure*}

The light curves of the optical and near-IR brightness, polarization degree, and polarization angle of PKS~2155$-$304 are plotted in Figure~\ref{detailed_opt_plot}, which is a zoomed-in version of Figure~\ref{ixpe_all}. The polarization properties are averaged within bins that are separated by the daily optical gaps. Additionally, to show the level of intra-night variability in the optical and near-IR bands, in Figure~\ref{detailed_opt_plot_nightly} we have zoomed-in (without performing any temporal rebinning) to three example nights with good time coverage, namely MJDs 60245, 60247, and 60248.

Figure~\ref{detailed_opt_plot} shows that, during the IXPE pointing, the optical brightness behavior was chromatic, with higher fluxes at longer-wavelength bands. Although this could be partly caused by the unknown host-galaxy-contributions, especially in case of the near-IR bands for which the host-contamination is expected to be most prominent, it is normal for the flux to be an increasing function of wavelength in blazars. Otherwise, we find that the optical and near-IR band polarization properties varied rather achromatically. However, the polarization degree at B band, $\Pi_\mathrm{O(B)}$, was consistently greater than that of the host-galaxy-corrected R-band, $\Pi_\mathrm{O(R^\dag)}$. Measuring $\frac{\Pi_\mathrm{O(B)}}{\Pi_\mathrm{O(R^\dag)}}$ for simultaneous data points and averaging its value results in $1.2\pm0.2$. To compare this ratio to a theoretical prediction, we use the turbulence-plus-shock model that theoretically predicts the average polarization degree to be <$\Pi$>~$\approx$~$0.75\sqrt{f^2_\mathrm{ord}+(1-f_\mathrm{ord})^2/N(\lambda)}$, where $f_\mathrm{ord}$ refers to the fraction of the magnetic field that is well ordered, and $N(\lambda)$ refers to the number of turbulent cells as a function of wavelength, $\lambda$, which can be estimated as $N(\lambda) \propto \lambda^{1/2}$ (\citealt{marscher2022_theoretical_estimate}). For the two optical bands, R (median wavelength of 640~nm) and B (450~nm), we find  $N_\mathrm{B}/N_\mathrm{R}$~$\approx$~0.8. Assuming $f_\mathrm{ord}$~$\approx$~0.05 and $N_\mathrm{R}$~$\approx$~1000 for the R-band, as estimated by \cite{marscher2022_theoretical_estimate}, we derive <$\Pi_\mathrm{O(R)}$>~$\approx$~4.4\%. In case of the B-band, where $N_\mathrm{B}$ is estimated to be around $1000 \times 0.8 = 800$, we obtain <$\Pi_\mathrm{O(R)}$>~$\approx$~4.5\%. Therefore, the theoretical model results in $\left< \frac{\Pi_\mathrm{O(B)}}{\Pi_\mathrm{O(R)}} \right>$~$\approx$~1.03. Although this is much smaller than the measured ratio, it is still compatible with the observed value within the estimated uncertainty. If such a discrepancy is shown to be statistically significant with future high-cadence optical polarization observations of HSP blazars, it would signal an underlying discrepancy with the turbulence-plus-shock model.

Figure~\ref{detailed_opt_plot_nightly} shows clear signs of intra-night variability in the optical band of PKS~2155$-$304. Although a detailed analysis of these is beyond the scope of this paper, such densely sampled optical polarization light curves can be instrumental in understanding the underlying mechanisms at play in blazar jets (e.g., \citealt{Marscher2021_optical_intra_night_variability}).

\end{appendix}

\end{document}